\documentclass[pdflatex, sn-mathphys-num, iicol]{sn-jnl}

\usepackage{graphicx}%
\usepackage{multirow}%
\usepackage{amsmath,amssymb,amsfonts}%
\usepackage{amsthm}%
\usepackage{mathrsfs}%
\usepackage[title]{appendix}%
\usepackage{xcolor}%
\usepackage{textcomp}%
\usepackage{manyfoot}%
\usepackage{booktabs}%
\usepackage{algorithm}%
\usepackage{algorithmicx}%
\usepackage{algpseudocode}%
\usepackage{listings}%
\usepackage{natbib}%
\usepackage{physics}%

\definecolor{wred}{HTML}{FC2855}
\definecolor{worange}{HTML}{FD8F2A}
\definecolor{wgreen}{HTML}{15A94E}
\definecolor{wblue}{HTML}{3D8EDD}
\definecolor{wpurple}{HTML}{DF6DF7}

\begin{document}
    \title[Quantum Advantage in Resource Estimation]{Quantum Advantage in Resource Estimation}

    \author[1]{\fnm{William A.} \sur{Simon}}

    \author[1]{\fnm{Peter J.} \sur{Love}}

    \affil*[1]{\orgdiv{Department of Physics and Astronomy}, \orgname{Tufts University}, \orgaddress{\street{574 Boston Avenue}, \city{Medford}, \postcode{02155}, \state{MA}, \country{USA}}}
    
    \abstract{
        \textbf{
        Quantum computing promises the ability to compute properties of quantum systems exponentially faster than classical computers.
        Quantum advantage is achieved when a practical problem is solved more efficiently on a quantum computer than on a classical computer.
        Demonstrating quantum advantage requires a powerful quantum computer with low error rates and an efficient quantum algorithm that has a useful application.
        Despite rapid progress in hardware development, we still lack useful applications that are feasible for the next generation of quantum computers.
        Here we argue that an exponential quantum advantage exists in producing numerical resource estimates of larger quantum algorithms by accurately measuring simulation errors.
        We provide a quantum algorithm for measuring simulation errors of Trotter-based algorithms.
        Our results indicate that this method will reduce runtimes of quantum algorithms by approximately three orders of magnitude for one-hundred qubit systems.
        We also predict that these reductions will increase with system size.
        The methods we propose require relatively few qubits and operations, meaning the next generation of quantum computers could compute simulation errors for classically intractable systems.  
        Since the underlying computations that lead to reduced resource estimates are infeasible for classical computers, this task is a candidate for demonstrating practical quantum advantage.
        }
    }

    \maketitle

    \section{Introduction}\label{sec:intro}

    \begin{figure*}
        \centering
        \includegraphics[width=\linewidth]{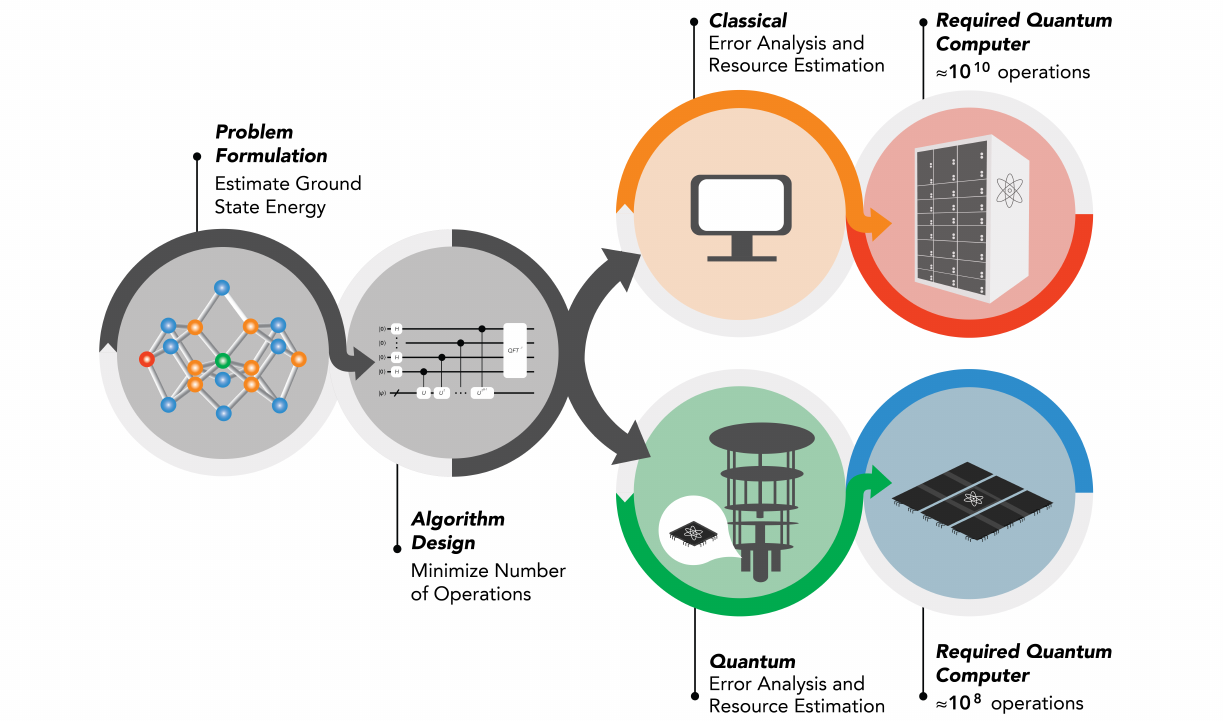}
        \caption{\textbf{Resource Estimation Workflow}
        Resource estimation begins by formulating the problem and defining the objective of the computation.
        An algorithm is chosen that performs the desired computation and an analysis of the error sources in the simulation is performed.
        The magnitude of the simulation error is estimated using either classical or quantum computers.
        The estimated simulation error results in an implementation of the algorithm with fixed quantum resources; these estimates determine the size of the quantum computer required to solve the problem.
        }
        \label{fig:schematic}
    \end{figure*}

    Quantum computers promise computational power exceeding that of classical computers \cite{benioff1980computer, feynman1982simulating}.
    Several quantum algorithms showing exponential reductions in computational resources over classical algorithms have been proposed, with useful applications ranging from quantum simulation \cite{feynman1982simulating, lloyd1996universal, qsimreview} and eigenvalue estimation \cite{abrams1999quantum} to factoring \cite{shor1999polynomial} and solving linear systems of equations \cite{harrow2009quantum}.

    The largest quantum computers in existence today have dozens to hundreds of physical qubits and are able to perform thousands of operations.
    These machines, which cannot perform fault-tolerant quantum computation, have been coined Noisy Intermediate-Scale Quantum (NISQ) devices \cite{preskill2018quantum}.
    NISQ devices have been used to run end-to-end quantum algorithms \cite{peruzzo2014variational, google2020hartree}, verify components of fault-tolerant quantum computation \cite{google2023suppressing, bluvstein2024logical, rodriguez2024experimental, lacroix2024scaling}, and most notably perform computations beyond classical tractability \cite{arute2019quantum, zhong2020quantum}.

    A long-awaited milestone is the experimental demonstration of ``quantum advantage": a computational speedup for a practically useful application that cannot be solved classically within a reasonable amount of time \cite{daley2022practical,lanes2025framework}.
    Despite several attempts \cite{kim2023evidence, king2025beyond}, NISQ devices have not been able to achieve quantum advantage \cite{tindall2024efficient, tindall2025dynamics, mauron2025challenging}.

    The next generation of quantum computers will have dozens to hundreds of \textit{logical} qubits, and aim to perform millions of operations.
    These devices are referred to as Early Fault-Tolerant Quantum Computers \cite{campbell2021early, katabarwa2024early} or MegaQuop machines \cite{preskill2025beyond}, and we will use the latter term in this work.

    It is currently unclear if MegaQuop machines will achieve quantum advantage.
    State of the art resource estimates predict the need for thousands of logical qubits and billions of operations for industrial applications, including simulating FeMoCo \cite{lee2021even, caesura2025faster} or breaking RSA encryption \cite{gidney2025factor}.

    One potential application of small quantum computers is to advance the development of larger quantum computers.
    Iyer et al. \cite{iyer2018small} argue that small quantum computers will be useful for testing and optimizing fault-tolerant protocols.
    Likewise, Kyaw et al. \cite{kyaw2021quantum} suggest designing sub-modules for larger devices on small quantum computers.

    We propose that small quantum computers could accurately estimate the resources needed for algorithms that would be run on larger devices (Figure \ref{fig:schematic}).
    Obtaining accurate resource estimates is itself a computationally challenging task; it requires evaluating quantum simulation errors, which are intractable for classical computers to compute accurately.
    We argue that simulation errors should be computed directly on quantum computers, providing a quantum advantage by reducing resource estimates.

    In this work, we outline a method to compute the simulation error of Trotter-based algorithms on small quantum computers.
    Our numerical results reduce the number of operations needed for Trotter-based Quantum Phase Estimation by two orders of magnitude for $14$ qubit systems, and predict a reduction of three orders of magnitude for $100$ qubit systems.
    Our resource estimates suggest that MegaQuop machines could provide quantum advantage in resource estimation by computing simulation errors of quantum algorithms.

    \section{Quantum Simulation Errors and Resource Estimation}
    \label{sec:qre}

    Quantum simulation algorithms aim to compute properties of quantum systems.
    One promising application in quantum simulation is estimating static properties, such as the energy levels of a Hamiltonian ($H$):
    \begin{equation}
        \label{eq:eigenvalue-equation}
        H \ket{\lambda_k} = \lambda_k \ket{\lambda_k}
    \end{equation}
    where $\lambda_k$ is the energy of the $k^\text{th}$ eigenstate ($\ket{\lambda_k}$).
    Quantum Phase Estimation (QPE) is a quantum algorithm that produces estimates of the eigenvalues of a Hamiltonian that are accurate up to a target error ($\epsilon$).

    ``Trotterization" \cite{lie1893theorie, trotter1958approximation, trotter1959product, suzuki1976generalized,suzuki1985decomposition, suzuki1986quantum, suzuki2012quantum, suzuki1990fractal, hatano2005finding} refers to several techniques for constructing quantum circuits that approximate the time evolution operator generated by the Hamiltonian ($U_H(t)$).
    An approximate time evolution operator can be viewed as an exact time evolution operator generated by an \textit{effective} Hamiltonian ($H^\prime)$.
    In the context of eigenvalue estimation, the simulation error arises as a distance between the target eigenvalue ($\lambda_k$) and the corresponding eigenvalue of the effective Hamiltonian ($\lambda_k^\prime$).
    We refer to this simulation error as the ``Trotter error" ($\epsilon_k^\text{TS}$).

    The resources required by a quantum simulation algorithm are dictated by the simulation error.
    An analytical expression for the number of operations ($O$) required to estimate an eigenvalue of a Hamiltonian using Trotter-based QPE can be obtained following Kivlichan et al. \cite{kivlichan2020improved}:
    \begin{equation}
        \label{eq:kivlichan-resource-estimates}
        O \geq \frac{\pi 3^{3/2}W_k^{1/2} L}{\epsilon^{3/2}} - 2L
    \end{equation} 
    where $L$ is the number of terms in the Hamiltonian and $W_k$ is a constant with respect to $t$ such that $\epsilon_k^\text{TS} \leq W_k t^2$.

    Numerical resource estimates depend on the magnitude of $\epsilon_k^\text{TS}$ (and $W_k$), yet this value is unknown and computing it directly is itself an eigenvalue problem.
    Prior works that produce numerical resource estimates for Trotter-based QPE \cite{kivlichan2020improved, campbell2021early} compute a state-independent upper-bound of $\epsilon_k^\text{TS}$ that is classically efficient to compute, referred to as the commutator bound \cite{childs2018toward, childs2021theory}:
    \begin{equation}
        \label{eq:classical-commutator-bounds}
        \epsilon_k^\text{TS} \leq W_C t^2
    \end{equation}
    where $W_C$ is constant with respect to $t$.
    This classically efficient algorithm upper-bounds the simulation error, resulting in compiled algorithms that require more resources than are strictly necessary.

    \section{Quantum Computing Simulation Error}
    \label{sec:algorithm}

    \begin{figure}
        \centering
        \includegraphics[width=\linewidth]{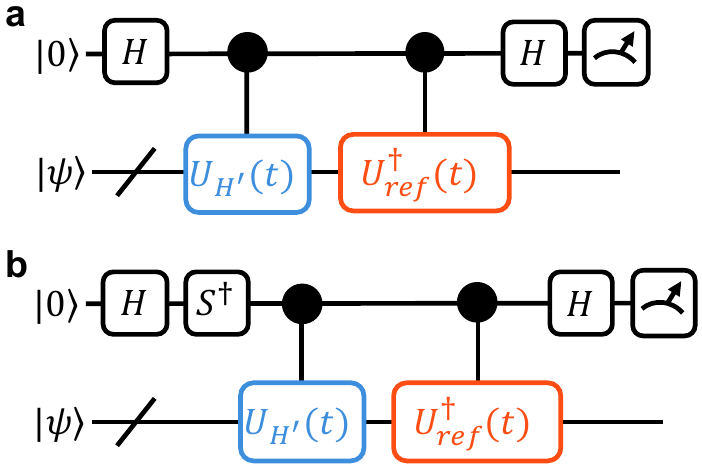}
        \caption{\textbf{Hadamard Test for Computing Phase Error} 
        A Hadamard test is a quantum algorithm for amplitude estimation. 
        \textbf{a,} The Hadamard test computing the real component of the amplitude, $\bra{\psi} U_\text{ref}^\dagger(t) U_{H^\prime}(t) \ket{\psi}$, is shown. 
        \textbf{b,} The circuit for evaluating the imaginary component of the amplitude is shown. 
        }
        \label{fig:trotter-had-test}
    \end{figure}

    The simulation error metric described by Eq. \ref{eq:classical-commutator-bounds} is independent of the quantum state upon which the algorithm acts.
    A state-dependent error metric that approximates $\epsilon_k^\text{TS}$ can be constructed from the phase error \cite{yi2022spectral}:
    \begin{equation}
        \label{eq:phase-error}
        \theta_\psi \equiv \arg(\bra{\psi} U_H^\dagger(t) U_{H^\prime}(t) \ket{\psi})
    \end{equation}
    where $U_{H^\prime}(t)$ is the time evolution of the effective Hamiltonian.
    The phase error on an eigenstate approximates the Trotter error on the corresponding eigenvalue:
    \begin{equation}
        \label{eq:phase-error-approximates-trotter-error}
        \epsilon_k^\text{TS} \approx t^{-1} |\theta_{\lambda_k}| \equiv \epsilon_k^\theta
    \end{equation}

    Amplitude estimation algorithms can compute an approximation to the phase error:
    \begin{equation}
        \label{eq:approximate-phase-error}
        \tilde{\theta}_\psi \equiv \arg(\bra{\psi}U_\text{ref}^\dagger(t) U_{H^\prime}(t)\ket{\psi})
    \end{equation}
    where we refer to $U_\text{ref}^\dagger(t)$ as the ``reference approximation''.
    The Hadamard test \cite{cleve1998quantum} (Figure \ref{fig:trotter-had-test}) is a quantum algorithm for amplitude estimation that requires one ancilla qubit and $\mathcal{O}(1/\delta^2)$ shots to estimate $\tilde{\theta}_\psi$ up to a precision of $\delta$.
    Other amplitude estimation algorithms require only $\mathcal{O}(1/\delta)$ shots at the expense of deeper quantum circuits \cite{brassard2000quantum,grinko2021iterative, suzuki2020amplitude}. 
    We opt for reducing the circuit depth to make this more feasible for NISQ and MegaQuop machines.

    The reference approximation in Eq. \ref{eq:approximate-phase-error} incurs an additional error term.
    This error term can lead to an overestimate or an underestimate of $\theta_\psi$.
    We propose using a ${p^\prime}^\text{th}$-order Suzuki-Trotter decomposition for the reference approximation, which allows us to systematically measure more accurate approximations by increasing $p^\prime$ as shown in Figure \ref{fig:phase_error_extrapolation}.

    The cost of computing $\tilde{\theta}_\psi$ is determined by the cost of implementing $U_\text{ref}^\dagger(t)$.
    Provided that an accurate reference approximation can be implemented with sufficiently low overhead, this gives an efficient quantum algorithm for measuring the simulation error of Trotter-based algorithms.

    \section{Results}
    \label{sec:quantum-qre}

    \begin{figure*}
        \centering
        \includegraphics[width=\linewidth]{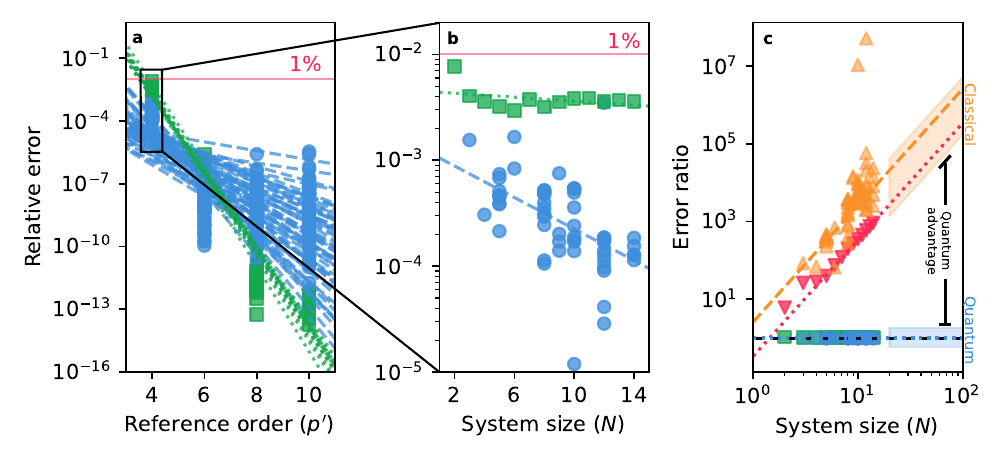}
        \caption{\textbf{Approximating Simulation Error}
        \textbf{a,}
        The relative error on the approximate phase error ($\tilde{\theta}_\psi$) is shown as a function of the Suzuki-Trotter order of the reference approximation ($p^\prime$).
        Results are shown for the ground state ($\epsilon_0^\text{TS}$) of chemistry Hamiltonians (blue circles) and random Pauli Hamiltonians (green squares).
        The target approximation is a second-order Suzuki-Trotter decomposition ($p=2$) with $t = \pi / ||4 H||$.
        The red line depicts a relative error of $1\%$.
        Two chemistry systems with $|\theta_\psi| < 10^{-10}$ are omitted since floating point errors obscure our numerics.
        \textbf{b,}
        The relative error on the approximate phase error ($\tilde{\theta}_\psi$) is shown as a function of system size ($N$).
        Results are shown for the ground state ($\epsilon_0^\text{TS}$) of chemistry Hamiltonians (blue circles) and random Pauli Hamiltonians (green squares).
        The target approximation is a second-order Suzuki-Trotter decomposition ($p=2$) and the reference approximation is a fourth-order Suzuki-Trotter decomposition ($p^\prime = 4$) with $t = \pi / ||4 H||$.
        The red line depicts a relative error of $1\%$.
        \textbf{c,} 
        The accuracy of the approximations to the Trotter error on the ground state ($\epsilon_0^\text{TS}$) is shown as a function of the system size ($N$).
        The error ratio is calculated as $|\epsilon^* / \epsilon_0^\text{TS}|$ where $\epsilon^*$ represents either the classical commutator bounds (Eq. \ref{eq:classical-commutator-bounds}) or the quantum phase error approximation (Eq. \ref{eq:phase-approximation-error}).
        The orange upward (red downward) triangles correspond to the classical error estimates and the blue circles (green squares) correspond to the quantum error estimates for chemistry systems (random Pauli Hamiltonians).
        The target approximation is a second-order Suzuki-Trotter decomposition ($p=2$) and the reference approximation is a fourth-order Suzuki-Trotter decomposition ($p^\prime = 4$) with $t = \pi / ||4 H||$.
        The dashed black line denotes an error ratio of $1$.
        }
        \label{fig:phase_error_extrapolation}
    \end{figure*}

    \begin{figure*}
        \centering
        \includegraphics[width=\linewidth]{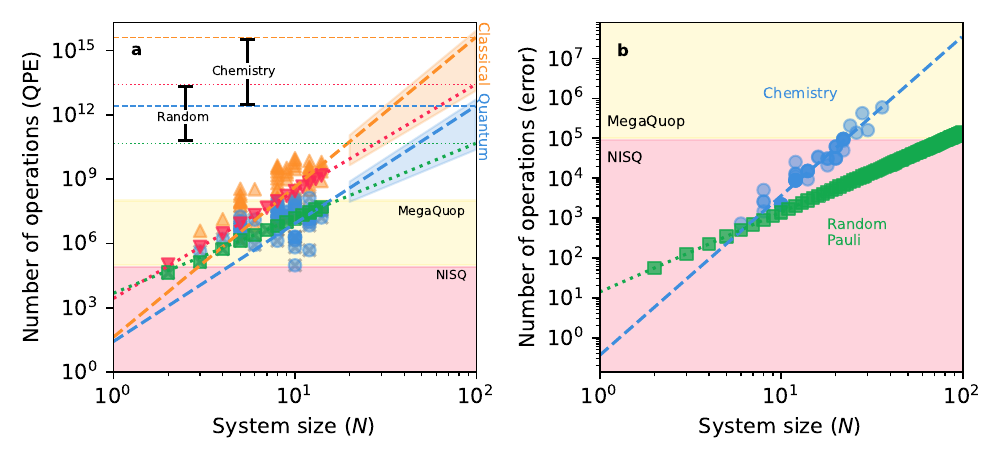}
        \caption{
            \textbf{Resource Estimates}
            \textbf{a,} The number of operations ($O$) required to compute the ground state energy using QPE is shown as a function of system size.
            The total error budget of QPE is set to $\epsilon = 0.0016$ Hartree ($\epsilon = 0.01$) for the for the chemistry systems (random Pauli Hamiltonians).
            The orange upward (red downward) triangles correspond to the resource estimates produced using classical computers and the blue circles (green squares) correspond to the resource estimates produced using quantum computers for chemistry systems (random Pauli Hamiltonians).
            The gray crosses correspond to resource estimates produced using the Trotter error on the ground state ($\epsilon_0^\text{TS}$). 
            The shaded yellow (red) regions depict the systems for which we could compute ground state energies on MegaQuop (NISQ) machines.
            We set the upper limit of MegaQuop (NISQ) machines at fewer than $10^8$ ($10^5$) operations.
            \textbf{b,} The number of operations required to compute the simulation error ($\tilde{\theta}_{\psi}$) is shown as a function of system size.
            These estimates assume the target approximation is a $p = 2$ Suzuki-Trotter decomposition and the reference approximation is a $p^\prime = 4$ Suzuki-Trotter decomposition.
            The blue circles (green squares) correspond to resource estimates for chemistry systems (random Pauli Hamiltonians).
            The shaded yellow (red) regions depict the systems for which we could compute the simulation error using MegaQuop (NISQ) machines.
            Operations are counted as the number of time evolutions of multi-qubit Pauli operators.
        }
        \label{fig:qres}
    \end{figure*}

    We benchmark our method on second-quantized Hamiltonians modeling the electronic structure of small molecules and random Pauli Hamiltonians.
    The chemistry Hamiltonians range from $3$ to $14$ qubits and contain between $34$ and $1819$ Pauli operators.
    The random Pauli Hamiltonians are defined by linear combinations of $N^2$ randomly sampled Pauli operators for systems ranging from $2$ to $14$ qubits.

    We calculate the relative error of $\tilde{\theta}_\psi$ for a second-order approximation ($p=2$) as a function of $p^\prime$ in subfigure \ref{fig:phase_error_extrapolation}a.
    The relative error decreases exponentially as a function of $p^\prime$ and is less than $1\%$ when $p^\prime = 4$ for all systems that we analyze.

    The dependence of the relative error on system size ($N$) is shown in subfigure \ref{fig:phase_error_extrapolation}b.
    We do not see a positive correlation between the relative error and $N$, suggesting that the resources required to compute $\tilde{\theta}_\psi$ will scale only with the number of operators in the Hamiltonian ($L$).

    We compare the ratio of both the commutator bounds (Eq. \ref{eq:classical-commutator-bounds}) and the phase error approximation:
    \begin{equation}
        \label{eq:phase-approximation-error}
        \epsilon_k^{\tilde{\theta}} \equiv t^{-1} | \tilde{\theta}_{\lambda_k} |
    \end{equation}
    to the Trotter error on the ground state ($\epsilon_0^\text{TS}$).
    These results are shown in subfigure \ref{fig:phase_error_extrapolation}c.
    The degree to which the commutator bounds overestimate the Trotter error grows with system size, while the approximate phase error on an eigenstate (Eq. \ref{eq:phase-approximation-error}) closely approximates the Trotter error regardless of system size.
    This indicates that classical computers will overestimate simulation errors for larger systems, while quantum computers can compute these simulation errors accurately. 

    Numerical resource estimates for computing ground state energies with QPE are shown in subfigure \ref{fig:qres}a.
    These results show that using a quantum computer to compute the simulation error reduces resource estimates by several orders of magnitude.
    We extrapolate from these trends to predict a reduction of approximately three orders of magnitude for $100$ qubit systems.
    These reductions make it possible to run QPE on small molecules using MegaQuop machines.

    Finally, we show the resource estimates for computing $\tilde{\theta}_\psi$ in subfigure \ref{fig:qres}b.
    We neglect the cost of state preparation, which could be significant when preparing eigenstates, however, computing the phase error on efficiently preparable states that are classically inaccessible would still be useful.
    These results show that NISQ and MegaQuop machines could compute simulation errors for $100$ qubit systems.

    \section{Conclusion and Outlook}
    \label{sec:conclusions}

    A core principle in quantum simulation is the expected computational efficiency of quantum computers for computing quantum properties.
    The simulation error associated with quantum algorithms is an inherently quantum metric, suggesting that quantum computers could calculate simulation errors efficiently.
    Accurate estimates of simulation errors lead to accurate resource estimates for larger quantum algorithms, implying that resource estimation is a natural application for quantum computers to provide quantum advantage.  

    We make this claim concrete by providing a method for computing the simulation error associated with Trotter-based algorithms on a quantum computer.
    One could extend these ideas to compute other simulation errors, such as the impact of rotation synthesis errors in a fault-tolerant architecture or the errors arising from approximating functions of the Hamiltonian using techniques such as QSVT \cite{gilyen2019quantum}.

    We provide resource estimates for quantum computing the simulation error of Trotter-based algorithms. 
    Proof-of-principle demonstrations of such error estimation on NISQ devices would yield reduced resource estimates for simulations that could be performed on MegaQuop machines.
    Resource estimation on MegaQuop machines is a candidate for quantum advantage; these calculations would yield reduced resource estimates for simulations that require the generation of fault-tolerant machines beyond the MegaQuop scale.

    The computational complexity of evaluating error metrics of quantum algorithms hints toward a synergy in the development cycle of quantum computers.
    Current-generation quantum computers can measure simulation errors efficiently, leading to a better understanding of the computational power of future generations of quantum computers.
    This echoes the development cycle of classical computers, where computer simulations are used to design future generations of computers.

    \section{Acknowledgments}
    We thank Sukin Sim, William M. Kirby, Carter M. Gustin, Alexis Ralli, and Feng Qian for comments and feedback on the manuscript.
    We thank Samantha Saltzman for the graphic design of Figure 1.
    W.A.S. is supported by the Department of Defense (DoD) through the National Defense Science \& Engineering Graduate (NDSEG) Fellowship Program.
    P.J.L. is supported by the NSF STAQ program under award NSF Phy- 1818914/232580 and by the NSF NVQL program under award NSF OSI-2410675/2531350.

    \appendix
    \onecolumn

    \section{Trotterization}
    \label{appendix:trotterization}

    Running quantum simulation algorithms on digital quantum computers requires representing the time evolution operator generated by the Hamiltonian as a quantum circuit.
    One strategy for constructing such circuits is to assume we can implement time evolution operators generated by terms within a particular operator basis and that we can write our Hamiltonian as a linear combination of these terms.
    Under these assumptions, we can compose a quantum circuit that approximates the time evolution of the Hamilonian as a product of the time evolutions of the terms in the linear combination \cite{lloyd1996universal}.

    There are a series of techniques for generating approximate time evolution operators as products of time evolutions over individual terms.
    These techniques are commonly referred to collectively as ``Trotterization" \cite{kivlichan2020improved,campbell2021early,poulin2014trotter, tang2021qubit,babbush2014adiabatic}, Lie-Trotter-Suzuki methods \cite{yi2022spectral}, Suzuki-Trotter or Trotter-Suzuki methods \cite{hatano2005finding, babbush2015chemical, hastings2014improving, wecker2014gate, poulin2014trotter,raeisi2012quantum}, Lie-Trotter methods \cite{yi2022spectral, childs2021theory}, or other similar variations \cite{nielsen2001quantum,childs2009limitations,hempel2018quantum}.

    In the following subsections, we split the techniques attributed to Lie, Trotter, and Suzuki into two batches: those that change the constant factor on the leading error term in the approximation and those that change the order in $t$ up to which the approximation is exact.
    We will refer to the former set of techniques as Lie-Trotter product formulae following primarily the works of Lie \cite{lie1893theorie} and Trotter \cite{trotter1958approximation, trotter1959product} and the latter as Suzuki-Trotter decompositions following the works of Suzuki et. al \cite{suzuki1976generalized,suzuki1985decomposition, suzuki1986quantum, suzuki2012quantum, suzuki1990fractal, hatano2005finding} and Trotter \cite{trotter1958approximation, trotter1959product}.

    In this work, we focus on Hamiltonians that can be written as a linear combination of Pauli operators:
    \begin{equation}
        \label{appendix-eq:lcu-paulis}
        H = \sum_{l=0}^{L-1} \alpha_l P_l
    \end{equation}
    where the coefficients ($\alpha_l$) are real-valued and the number of operators ($L$) is polynomial in the system size ($N$).
    We define the system size ($N$) as the number of qubits upon which the Hamiltonian acts.
    The time evolution operators over individual Pauli operators can be efficiently compiled using a standard gateset \cite{nielsen2001quantum}, meaning these approximations could be run efficiently on hardware.

    \subsection{Lie-Trotter Product Formulae}
    \label{appendix-subsec:lie-trotter}

    For Hamilitonians described by Eq. \ref{appendix-eq:lcu-paulis}, the Lie-Trotter product formulae provide an approximation to the time evolution operator by breaking the total time evolution into slices of smaller time-steps:
    \begin{equation}
    \label{appendix-eq:lie-trotter-lcu}
        e^{-iHt} = \lim_{S \rightarrow \infty} \Big[\prod_{l=0}^{L-1} e^{-i \alpha_l P_l t/S}\Big]^S
    \end{equation}
    where $t$ is the total desired evolution time, which is achieved by $S$ consecutive approximate time evolutions of duration $t/S$.

    For an integer value of $S$, Eq. \ref{appendix-eq:lie-trotter-lcu} gives an approximation to the time evolution operator.
    We refer to the value of $S$ as the number of Trotter steps where each Trotter step evolves the system for duration $t/S$.
    Increasing the number of Trotter steps from $1$ to $S$ leads to a constant factor reduction in the error of the approximation by a factor of $S$.
    If we assume that the cost of implementing a time evolution operator of an individual term in the Hamiltonian is independent of $t$, then increasing the number of Trotter steps in a Lie-Trotter approximation from $1$ to $S$ increases the cost of implementing the approximation by a factor of $S$.

    The effective Hamiltonian for a given approximate time evolution operator is a function of the evolution time.
    The Lie-Trotter product formulae provide a means to implement the time evolution operator of the \textit{same} effective Hamiltonian for durations of $St$ when $S$ is a positive integer.
    Application of the unitary $U_{H^\prime}(St) = e^{-iH^\prime St}$ can be accomplished using $S$ Trotter steps that each implement an approximate time evolution for duration $t$.

    In the context of QPE, this ability to retain the same effective Hamiltonian throughout the algorithm allows for a more accurate analysis of the contribution of the simulation error induced by Trotterization to the overall error of the algorithm.
    Without this ability, it would not be valid to view the measurement outcomes of QPE as estimates of the eigenvalues of the effective Hamiltonian.

    \subsection{Suzuki-Trotter Decompositions}

    Suzuki  \cite{suzuki1976generalized, suzuki1990fractal} defined a $p^{\text{th}}$-order approximation to the time evolution operator as an approximation that agrees up to and including terms that are $p^{\text{th}}$-order in $t$.
    Following this definition, the Lie-Trotter product formula given by Eq. \ref{appendix-eq:lie-trotter-lcu} \cite{lie1893theorie, trotter1958approximation, trotter1959product} results in a first-order approximation ($p=1$) for a Hamiltonian of the form in Eq. \ref{appendix-eq:lcu-paulis}:
    \begin{equation}
    \label{appendix-eq:first-order-trotter-lcu}
        U_H(t) \approx U^\prime_{1}(t) = \prod_{l=0}^{L-1} e^{-i \alpha_l P_l t}
    \end{equation}
    where the $*_1$ subscript indicates a first-order approximation.
    The value of $S$ does not change the order in $t$ to which the approximation is exact, therefore we implicitly set $S=1$ in the above equation.  
    In this work, we will refer to Eq. \ref{appendix-eq:first-order-trotter-lcu} as the first-order Suzuki-Trotter decomposition.

    A second-order approximation for a Hamiltonian of the form of Eq. \ref{appendix-eq:lcu-paulis} in is given by \cite{suzuki1986quantum, suzuki2012quantum, suzuki1985decomposition}:
    \begin{equation}
        \label{appendix-eq:second-order-trotter-lcu}
        U^\prime_{2}(t) = \big( \prod_{l=0}^{L-1} e^{-i \alpha_l P_l \frac{t}{2}} \big) \big( \prod_{l={L-1}}^{0} e^{-i \alpha_l P_l \frac{t}{2}} \big)
    \end{equation}
    which we will refer to as the second-order Suzuki-Trotter decomposition.
    In the second-order Suzuki-Trotter decomposition, each term is time-evolved over one half of the desired evolution time until all terms have been time-evolved.
    Then, each term is time-evolved again, but the order of the terms is reversed.

    Suzuki also provided higher-order decompositions for even values of $p$, which are defined recursively based on the second-order decomposition:
    \begin{equation}
        \label{appendix-eq:higher-order-trotter-decompositions}
        \begin{split}
            U^\prime_p(t) &= \Big[ U^\prime_{p^*}(\alpha_p t) \Big]^2 U^\prime_{p^*}\big((1 - 4\alpha_p)t\big) \Big[ U^\prime_{p^*}(\alpha_p t) \Big]^2 \\
            \alpha_p &= (4 - 4^{1 / (p-1)})^{-1}
        \end{split}
    \end{equation}
    where $p^* = p - 2$ \cite{suzuki1990fractal}.
    We will refer to these approximations as $p^\text{th}$ order Suzuki-Trotter decompositions.

    Due to this recursive construction, we can measure the cost of a $p^\text{th}$-order Suzuki-Trotter decomposition by counting how many times the second-order Suzuki-Trotter decomposition is called:
    \begin{equation}
        \label{appendix-eq:query-complexity-higher-order-trotter}
        \mathcal{Q}_p = 5^{(p/2)-1}
    \end{equation}
    where $\mathcal{Q}_p$ is the number of queries to the second-order Suzuki-Trotter decomposition required to implement a $p^\text{th}$-order Suzuki-Trotter decomposition with $S=1$.

    The effective Hamiltonian is dependent on the order of the Suzuki-Trotter decomposition.
    Therefore the same Suzuki-Trotter decomposition must be used throughout an algorithm if the intention is to retain the same effective Hamiltonian.

    All Trotter-based methods specify an ordering of the time evolutions of the individual terms in Eqs. \ref{appendix-eq:lie-trotter-lcu}, \ref{appendix-eq:first-order-trotter-lcu}, and \ref{appendix-eq:second-order-trotter-lcu}.
    Different orderings of these terms will result in different approximations and subsequently different effective Hamiltonians \cite{tranter2019ordering}. 
    Many works focus on finding an ordering of the terms that increases the accuracy of the approximation for specific Hamiltonians \cite{kivlichan2020improved, campbell2021early}.

    \subsection{Trotter Error}

    The term ``Trotter Error" is used throughout the literature to refer to several simulation errors that arise when using Trotterization to approximate the time evolution of a linear combination of operators.

    Many works define ``Trotter error" as the distance between the time evolution operator and the approximate time evolution operator and quantify this distance using the operator norm: $|| U(t) - U^\prime(t) ||$.
    In this context, the operator norm ($||O||$) refers to the spectral norm, which is given by \cite{childs2009limitations}: 
    \begin{equation}
        \label{methods-eq:spectral-norm}
        \max_{||\ket{\psi}||_2 = 1} || O \ket{\psi} ||_2
    \end{equation}
    where $||*||_2$ denotes the L2-norm (also referred to as the Euclidean norm) of the vector.
    The spectral norm of an operator can be equivalently defined as the largest singular value of the operator.

    Other works that focus on the state-dependent ``Trotter error" \cite{yi2022spectral, burgarth2024strong} use the Euclidean distance between the desired time-evolved state and the achieved state: $|| (U(t) - U^\prime(t)) \ket{\psi} ||_2$.

    In this work, we define the ``Trotter Error'' ($\epsilon^\text{TS}_{k}$) on an eigenstate ($\ket{\lambda_k}$) as the magnitude of the distance between the target eigenvalue ($\lambda_k$) and the corresponding eigenvalue of the effective Hamiltonian ($\lambda^\prime_k$): 
    \begin{equation}
        \label{methods-eq:trotter-error}
        \epsilon^\text{TS}_{k} \equiv |\lambda_k - \lambda^\prime_k|
    \end{equation}
    This is a slight generalization of the definition used by Babbush et al. \cite{babbush2015chemical}, which assumes the target eigenstate is the ground state.

    To compute the effective eigenvalues and effective eigenstates of a given Trotterization, we first generate the matrix form of the effective time evolution operator.
    This matrix and the matrix representing the Hamiltonian are diagonalized to find the respective eigenvalues and eigenstates.
    Since these two Hamiltonians are not guaranteed to share the same eigenbasis, there is an ambiguity over the correspondance between the eigenstates of the Hamiltonian and the eigenstates of the effective Hamiltonian.
    We assign the effective eigenvector that has the largest squared overlap with the target eigenstate as the corresponding effective eigenvector.
    The effective eigenvalue of the effective Hamiltonian is computed by taking the angle of the corresponding eigenvalue of the effective time evolution unitary and dividing by the evolution time.

    \section{Resource Estimation: QPE}
    \label{appendix:minimizing-qres}

    \begin{figure*}
    \begin{center}
            \centering
            \includegraphics[width=\linewidth]{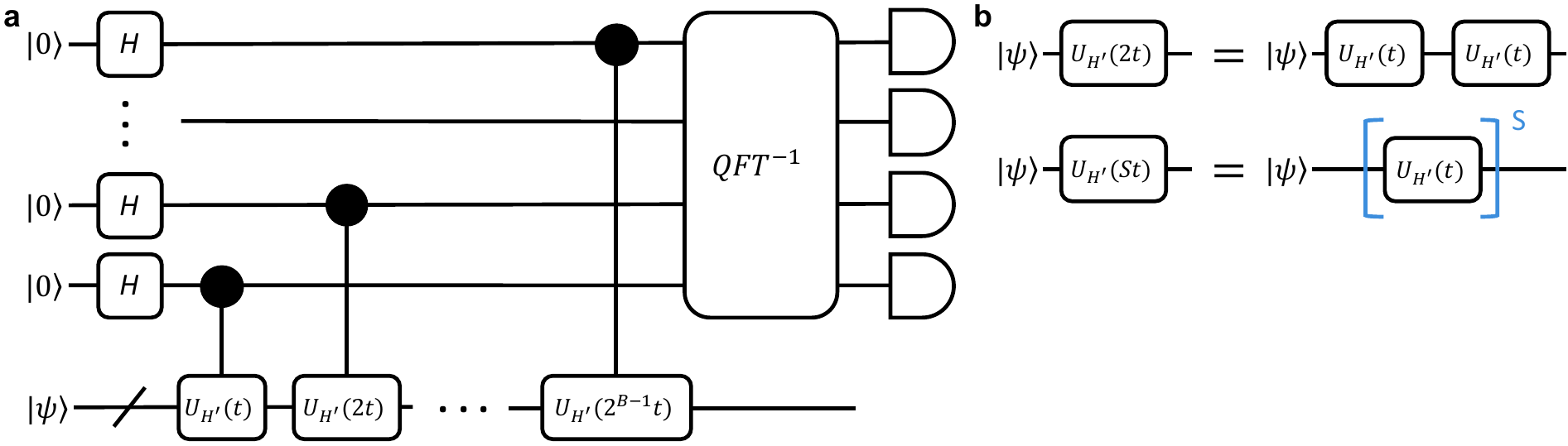}
            \caption{
                \textbf{Trotter-Based Quantum Phase Estimation} 
            \textbf{a,} A quantum circuit for QPE using an approximate time evolution unitary is shown.
            We refer to the qubits beginning in the $\ket{0}$ state as the phase register and the number of these qubits is given by $B$.
            \textbf{b,}
            Applying effective time evolution operations of durations $St$ is achieved by $S$ applications of the effective time evolution unitary for time $t$.
            }
            \label{extended-fig:qpe-circuit}
    \end{center}
    \end{figure*}

    Here, we review the resource estimation protocol for Trotter-based QPE used by Kivlichan et. al \cite{kivlichan2020improved} and Campbell \cite{campbell2021early}.
    We adapt this routine assuming the use of the canonical QPE circuit (Figure \ref{extended-fig:qpe-circuit}) as opposed to the iterative QPE algorithm \cite{higgins2007entanglement,berry2009,wecker2014gate} assumed by Kivlichan et al. and Campbell.
    The spacetime quantum resources are quantified by the number of qubits (space) and the number of operations (time) required to run a quantum algorithm.
    These resource estimates are obtained by analytically solving for the evolution time of the base unitary in QPE that minimizes the number of calls to the approximate time evolution unitary, subject to the constraint that the error on the estimate of the target eigenvalue is less than the overall error budget.

    We assume that the overall error on the estimate of the eigenvalue obtained by QPE is dominated by the Trotter error (Eq. \ref{methods-eq:trotter-error}) and the discretization error of QPE:
    \begin{equation}
        \label{appendix-eq:discretization-error}
        \epsilon^{QPE} \leq \pm \frac{\pi}{t 2^{B}}
    \end{equation}
    where $\epsilon^{QPE}$ is quantified in the energy-scale of the Hamiltonian.

    We assume that the repeated calls to the approximate time evolution unitary dominant the number of operations in the algorithm.
    We will neglect the number operations required for state preparation, as well as the operations required for the inverse quantum fourier transform and measurement and readout at the end of QPE. 

    First, we focus on a second-order Suzuki-Trotter decomposition using one Trotter step as this allows us to compare directly with resource estimates obtained using the commutator bounds (Eq. \ref{eq:classical-commutator-bounds}).
    Then, we generalize these results to higher-order Suzuki-Trotter decompositions ($p > 2$) and multiple Trotter steps ($S > 1$).

    We quantify the required resources in terms of the number of time evolution operations ($O$) of multi-qubit Pauli operators.
    For a given Hamiltonian, this can be computed as the number of queries ($Q$) to the second-order Suzuki-Trotter decomposition multiplied by two times the number of terms in the Hamiltonian ($2L$).
    Since the number of qubits needed in the phase register ($B$) determines the number of queries ($Q$), minimizing either quantity ($B$ or $Q$) will suffice to minimize both the number of qubits and the number of operations.

    In total, to obtain a single measurement from QPE in the form of Figure \ref{extended-fig:qpe-circuit} will require the following number of queries to the approximate time evolution unitary:
    \begin{equation}
        \label{appendix-eq:query-complexity}
        Q = 2^B - 1
    \end{equation}

    In the energy scale of the Hamiltonian, the error constraint for a second-order Suzuki-Trotter decomposition is given by:
    \begin{equation}
        \label{appendix-eq:error-constraint-second-order-trotter}
        \epsilon^{QPE} + \epsilon^\text{TS}_k = \frac{\pi}{t 2^{B}} + W_kt^2 \leq \epsilon
    \end{equation}
    which directly implies: 
    \begin{equation}
        \label{appendix-eq:b-in-terms-of-t}
        B \geq \log_2 \Big[ \frac{\pi}{t(\epsilon - W_kt^2)} \Big]
    \end{equation}
    \begin{equation}
        \label{appendix-eq:q-in-terms-of-t}
        Q \geq \frac{\pi}{t(\epsilon - W_kt^2)} - 1
    \end{equation}

    To find the value of $t$ that minimizes $Q$, we take the partial derivative of $Q$ with respect to $t$: 
    \begin{equation}
    \label{appendix-eq:optimization-of-B}
        \begin{split}
        0 &= \frac{\partial}{\partial t} Q(t) = \frac{\partial}{\partial t} \Big[ \frac{\pi}{t(\epsilon - W_kt^2)} - 1 \Big] = \frac{\partial}{\partial t} \Big[ \frac{\pi}{t(\epsilon - Wt^2)} - 1 \Big] \\ 
        &= \pi \Big[ -t^{-2} (\epsilon - W_kt^2)^{-1} + (2W_kt) t^{-1} (\epsilon - W_kt^2)^{-2} \Big] \\
        &= -t^{-2} (\epsilon - W_kt^2)^{-1} + (2W_kt) t^{-1} (\epsilon - W_kt^2)^{-2}
        \end{split}
    \end{equation}
    and then solve for the value of $t$ at the critical points:
    \begin{equation}
        \label{appendix-eq:appendix-optimal-time-commutator}
        \begin{split}
            t_*^{-2}(\epsilon - W_kt_*^2)^{-1} &= 2W_k (\epsilon - W_kt_*^2)^{-2} \\
            (\epsilon - W_kt_*^2)^{2} &= 2W_kt_*^2(\epsilon - W_kt_*^2) \\
            \epsilon - W_kt_*^2 &= 2W_kt_*^2 \\
            t_* &= \sqrt{\frac{\epsilon}{3W_k}}
        \end{split}
    \end{equation}
    where $t_*$ is the value of $t$ that minimizes $Q$.

    Substituting $t_*$ into Eq. \ref{appendix-eq:error-constraint-second-order-trotter} leads to:
    \begin{equation}
        \begin{split}
        \epsilon^\text{QPE} \approx \frac{2\epsilon}{3} \\
        \epsilon^\text{TS} \approx \frac{\epsilon}{3} 
        \end{split}
    \end{equation}
    implying that the optimal split between the error sources occurs when two-thirds of the error budget is allocated to the discretization error of QPE and one-third of the error budget is allocated to the Trotter error.

    It should be noted that if $t_*$ exceeds the value:
    \begin{equation}
        \label{appendix-eq:t-max}
        t_\text{max} = \frac{\pi}{||H^\prime||}
    \end{equation}
    where $||H^\prime||$ is the spectral norm of the effective Hamiltonian, then the effective eigenphases are not guaranteed to fit within the range $[-\pi, \pi)$.
    Since the estimates obtained from QPE are measured as $\lambda^\prime_k t \mod 2\pi$, this could create an ambiguity in the measurement results obtained from QPE.

    To ensure that this ambiguity on the value of $\lambda^\prime_k$ does not exist, $t_*$ must be set to be less than or equal to $t_\text{max}$.
    For most relevant systems, it is expected that $||H^\prime||$ grows with system size and hence we expect $t_\text{max} \ll 1$ asymptotically.
    In this work, we do not explicitly restrict $t_* \leq t_\text{max}$ for resource estimates of QPE since $t_*$ is a function of the desired error budget ($\epsilon$).

    Substituting $t_*$ into Eqs. \ref{appendix-eq:b-in-terms-of-t} and \ref{appendix-eq:q-in-terms-of-t} results in the optimal values of $B$, $Q$, and $O$:
    \begin{equation}
        \label{appendix-eq:optimal-bop}
        B_* \geq \log_2 \Big[ \frac{\pi 3^{3/2} W_k^{1/2}}{2 \epsilon^{3/2}} \Big]
    \end{equation}
    \begin{equation}
        \label{appendix-eq:optimal-query-complexity}
        Q_* \geq \frac{\pi 3^{3/2} W_k^{1/2}}{2 \epsilon^{3/2}} - 1
    \end{equation}
    \begin{equation}
        \label{appendix-eq:optimal-num-ops}
        O_* \geq \frac{\pi 3^{3/2} W_k^{1/2}L}{\epsilon^{3/2}} - 2L
    \end{equation}

    The choice of the error metric used to approximate the Trotter error constant ($W_k$) will lead to different values of the number of qubits in the phase register and the number of calls to the approximate time evolution unitary.
    For example, using the error constant associated with the phase error ($W_{\theta, k}$) as opposed to the commutator bounds ($W_C$) will lead to a reduction in the in the number of qubits required in the phase register of $\log_2 \sqrt{W_C / W_{\theta, k}}$ and a reduction in the number of operations by a factor of $\sqrt{W_C / W_{\theta, k}}$.

    \subsection{Higher-Order Suzuki-Trotter}

    Here, we generalize the above resource estimation protocol to account for higher-order Suzuki-Trotter decompositions ($p \geq 2$).
    One could apply similar methods as those shown here to minimize the resource estimates for a first-order Suzuki-Trotter decomposition ($p = 1$).

    The error constraint for an arbitrary-order Suzuki-Trotter decomposition is given by:
    \begin{equation}
        \label{appendix-eq:higher-order-qpe-error-constraint}
        \frac{\pi}{t 2^{B}} + W_{k, p} t^p \leq \epsilon
    \end{equation}
    where the $*_p$ subscript indicates that the constant $W$ is different for different values of $p$.
    As before, we assume that the value of $W_{k, p}$ is constant in the range of evolution times we consider.

    Following the recursive construction for higher-order Suzuki-Trotter decompositions (Eq. \ref{appendix-eq:higher-order-trotter-decompositions}), the number of calls to the second-order Suzuki-Trotter decomposition in QPE is given by:
    \begin{equation}
        Q_p = 5^{(p/2) - 1}(2^B - 1)
    \end{equation}

    Writing $Q_p$ in terms of $t$ using the above error-constraint (Eq. \ref{appendix-eq:higher-order-qpe-error-constraint}) leads to:
    \begin{equation}
        \label{appendix-eq:query-complexity-general-p}
        Q_p \geq 5^{(p/2) - 1}  \big( \frac{\pi}{t(\epsilon - W_{k, p} t^2)} - 1 \big)
    \end{equation}

    Taking the partial derivative with respect to $t$ and solving for $t_*$ results in:
    \begin{equation}
        t_* = \big( \frac{\epsilon}{(p+1)W_{k, p}} \big)^{-p}
    \end{equation}
    When $p = 2$, we recover the same form as Eq. \ref{appendix-eq:appendix-optimal-time-commutator}.

    Plugging $t_*$ into Eq. \ref{appendix-eq:query-complexity-general-p} gives:
    \begin{equation}
        \label{appendix-eq:higher-order-query-complexity}
        Q_p \geq 5^{(p/2) - 1}  \Big( \frac{\pi W_{k, p}^{1/p} (p + 1)^{(1/p) + 1} }{p \epsilon^{(1/p) + 1} } - 1 \Big)
    \end{equation}

    Since $W_{k, p}$ is different for different values of $p$, it cannot be immediately determined which value of $p$ will be optimal.
    Methods that estimate $W_{k, p}$ for higher-order decompositions are required to produce numerical resource estimates to allow for a direct comparison.

    \subsection{Multiple Trotter Steps}

    So far we have assumed that the base unitary in QPE is one Trotter step ($S = 1$).
    Here, we show that $S = 1$ minimizes the query complexity.
    This may make intuitive sense as using a value of $S$ that is a power of two would be equivalent to bit-shifting QPE.
    For simplicity, we assume a second-order Suzuki-Trotter decomposition.

    If the total evolution time of the base unitary in QPE is $St$, where each Trotter step evolves the system for time $t$, then each Trotter step accumulates an error term of order $\mathcal{O}(W_kt^2S^{-2})$.
    In total, the error term for the base unitary using $S$ Trotter steps becomes $\mathcal{O}(W_k t^2S^{-1})$. 
    As a result, the error constraint for QPE becomes:
    \begin{equation}
        \label{appendix-eq:error-constraint-multiple-trotter-steps}
        \frac{\pi}{St 2^{B}} + W_kt^2S^{-1} \leq \epsilon
    \end{equation}

    The number of calls to the second-order Suzuki-Trotter unitary in QPE is given by:
    \begin{equation}
        Q_S = S (2^B - 1)
    \end{equation}

    Writing $Q_S$ in terms of $t$ using the above error-constraint (Eq. \ref{appendix-eq:error-constraint-multiple-trotter-steps}) leads to:
    \begin{equation}
        \label{appendix-eq:query-complexity-general-s}
        Q_S = S \Big( \frac{\pi}{St (\epsilon - W_kt^2S^{-1})} - 1 \Big)
    \end{equation}

    Then, taking the partial derivative with respect to $t$ and solving for $t_*$ results in:
    \begin{equation}
        t_* = \sqrt{\frac{\epsilon S}{3W_k}}
    \end{equation}
    When $S = 1$, we recover the same form as Eq. \ref{appendix-eq:appendix-optimal-time-commutator}.

    Plugging $t_*$ into Eq. \ref{appendix-eq:query-complexity-general-s} gives:
    \begin{equation}
        Q_S \geq \frac{\pi 3^{3/2} W_k^{1/2}}{2 \epsilon^{3/2} S^{1/2}} - S
    \end{equation}

    The inequality $Q_S < Q_1$ will indicate which values of $S$ lead to requiring fewer operations:
    \begin{equation}
        \label{appendix-eq:inequality-of-S}
        \begin{split}
            Q_S &< Q_1 \\
            \frac{\pi 3^{3/2} W_k^{1/2}}{2 \epsilon^{3/2} S^{1/2}} - S &< \frac{\pi 3^{3/2} W_k^{1/2}}{2 \epsilon^{3/2} } - 1 \\
            \big[\frac{\pi 3^{3/2} W_k^{1/2}}{2 \epsilon^{3/2}}\big] &< (S - 1)(S^{-1/2} - 1)^{-1} \\
            \big[\frac{\pi 3^{3/2} W_k^{1/2}}{2 \epsilon^{3/2}}\big] &< - S - S^{1/2} \\
        \end{split}
    \end{equation}
    The values of $\epsilon$, $W_k$, and $S$ are all defined to be positive, real numbers.
    Therefore the left-hand side of Eq. \ref{appendix-eq:inequality-of-S} will be a positive number and the right-hand side will be a negative number.
    Consequently, no value of $S$ will satisfy the inequality, implying that $S = 1$ is optimal.

    \section{Commutator Bounds}
    \label{appendix:commutator_bounds}

    Trotter approximations admit an error that scales with the evolution time.
    The operator norm of the distance between the time evolution unitary and a $p^{th}$-order Suzuki-Trotter decomposition results in the following error term:
    \begin{equation}
        \label{appendix-eq:operator-norm}
        \Delta_p \equiv || U(t) - U^\prime_{p}(t) || \leq \mathcal{O}(t^{p+1})
    \end{equation}
    where $|| * ||$ denotes the operator norm (Eq. \ref{methods-eq:spectral-norm}).

    Following Bhatia and Davis \cite{bhatia1984bound}, Eq. \ref{appendix-eq:operator-norm} gives an upper-bound of the difference between the eigenvalues of the time evolution unitary and the approximate time evolution unitary:
    \begin{equation}
        \label{appendix-eq:op-norm-spectral-error}
        | e^{-i\lambda_k t} - e^{-i\lambda^\prime_k t} | \leq \Delta_p \leq \mathcal{O}(t^{p+1})
    \end{equation}

    Following Eq. 58 of Kivlichan et al. \cite{kivlichan2020improved}, the distance between the phases of the time evolution unitary ($\lambda_k t$) and the approximate time evolution unitary ($\lambda_k^\prime t$) can be upper bounded by:
    \begin{equation}
        \label{appendix-eq:effective-evalue-difference}
        | \lambda_k t - \lambda_k^\prime t | \leq \tan^{-1}(\Delta_p\frac{\sqrt{4 - \Delta_p^2}}{2 - \Delta_p^2}) = \Delta_p + \frac{\Delta_p^3}{24} + \mathcal{O}(\Delta_p^5)
    \end{equation}

    Under the assumption that the error contribution is dominated by the first-order in $\Delta_p$ (or that $| \lambda_k t - \lambda_k^\prime t| \ll 1$), Eqs. \ref{appendix-eq:op-norm-spectral-error} and \ref{appendix-eq:effective-evalue-difference} imply that the scaling of the distance between the eigenvalues of the Hamiltonian and eigenvalues of the effective Hamiltonian is given by:
    \begin{equation}
        \label{appendix-eq:trotter-error-scaling}
        \epsilon^\text{TS}_k \equiv | \lambda_k - \lambda_k^\prime | \leq t^{-1} \Delta_p \leq \mathcal{O}(t^p) = W_O t^p
    \end{equation}
    where $W_O$ is constant with respect to $t$ and is associated with the operator norm defined in Eq. \ref{appendix-eq:operator-norm}.

    Since Eq. \ref{appendix-eq:trotter-error-scaling} is state-independent, it results in an upper-bound on the maximum Trotter error over all eigenvalues:
    \begin{equation}
        \max_k \epsilon^\text{TS}_k \leq W_O t^p
    \end{equation}

    Computing either Eq. \ref{appendix-eq:op-norm-spectral-error} or Eq. \ref{appendix-eq:trotter-error-scaling} is itself an eigenvalue problem and therefore inefficient to compute classically.
    The commutator bounds give classically efficient upper-bounds on this operator norm for second-order Suzuki-Trotter decompositions.

    The commutator bounds are derived from the Baker-Campbell-Hausdorff relations \cite{baker1905alternants, hausdorff1906symbolische}.
    Prior works \cite{wecker2014gate,kivlichan2020improved, childs2021theory,campbell2021early} analyzing these commutator bounds lead to the following upper-bound of the spectral norm in Eq. \ref{appendix-eq:trotter-error-scaling} for a second-order Suzuki-Trotter decomposition:
    \begin{equation}
    \label{appendix-eq:second-order-commutator-bounds}
        || U(t) - U^\prime_2(t) || \leq \frac{t^3}{12} \sum_{b=1}^{M-1} \Big( || \sum_{c > b} \sum_{a > b} [[H_b, H_c], H_a] || + \frac{1}{2} || \sum_{c > b} [[H_b, H_c], H_b] || \Big)
    \end{equation}
    where $||*||$ again denotes the operator norm and $[A, B]$ denotes the commutator between operators $A$ and $B$.
    The operators $H_a$, $H_b$, and $H_c$ index the terms in Eq. \ref{appendix-eq:lcu-paulis}.

    The operator norm of this sum of nested commutators gives a constant with respect to the evolution time ($t$):
    \begin{equation}
    \label{appendix-eq:trotter-error-constant}
        W_C = \frac{1}{12} \sum_{b=1}^{M-1} \Big( || \sum_{c > b} \sum_{a > b} [[H_b, H_c], H_a] || + \frac{1}{2} || \sum_{c > b} [[H_b, H_c], H_b] || \Big)
    \end{equation}
    where $W_C \geq W_O$.

    In general, the cost of computing the operator norms in Eq. \ref{appendix-eq:trotter-error-constant} will scale exponentially with the locality of the operators in Eq. \ref{appendix-eq:lcu-paulis}.
    Some works that compute these commutator error constants ($W_C$) upper-bound these operator norms using additional information regarding the structure of the Hamiltonian \cite{campbell2021early}.

    The commutators of Pauli operators will themselves be Pauli operators and the operator norm of all Pauli operators is $1$.
    Therefore, the operator norm of a sum of commutators of Pauli operators can be bounded by the L1-norm of the coefficients of the commutators.
    The numerical values of $W_C$ that we report in this work are computed via this approach.

    \section{Phase Error}
    \label{appendix:phase_error}

    Here, we review the phase error ($\theta_\psi$) as defined by  Yi and Crosson \cite{yi2022spectral}.
    We then explicitly show how the phase error on an eigenstate provides an approximation to the Trotter error.

    \subsection{Definition}

    Yi and Crosson quantify the error incurred by an approximate time evolution unitary on a quantum state by:
    \begin{equation}
        \label{appendix-eq:trotter-effects}
        U^\prime(t) \ket{\psi} = \sqrt{1 - f_\psi} e^{i \theta_\psi} \ket{\psi (t)} + \sqrt{f} \ket{\psi^\perp(t)}
    \end{equation} 
    where $\ket{\psi(t)} \equiv U(t) \ket{\psi}$ is the desired time-evolved state and $\ket{\psi^\perp(t)}$ is a quantum state orthogonal to the desired state.

    The fidelity error, $f_\psi$, captures the infidelity between the achieved state and the desired state:
    \begin{equation}
        \label{appendix-eq:fidelity-error}
        f_\psi = 1 - | \bra{\psi} U^\dagger(t) U^\prime(t) \ket{\psi} |^2
    \end{equation}
    The fidelity error takes the value 0 when the approximation is exact and takes the value 1 in the worst case: when the achieved state is orthogonal to the desired state.
    Yi and Crosson note that the fidelity error corresponds to a reduction in the probability of observing estimates of the target eigenvalue in QPE.

    The phase error, $\theta_\psi$, captures the difference in phase between the achieved state and the desired state.
    The phase error takes the value 0 when the approximation is exact (no difference in phase between the acheived state and the desired state) and takes the value $\frac{\pi}{2}$ in the worst-case.

    One of the findings of Yi and Crosson is that the Euclidean distance between the achieved state and the desired state for an approximate time evolution operator can be upper and lower bounded by the phase error and the fidelity error:
    \begin{equation}
        f + \frac{\theta^2}{4} \leq || (U(t) - U^\prime(t)) \ket{\psi} ||^2 \leq 2f + \theta^2
    \end{equation}
    where $||*||$ denotes the Euclidean distance between the achieved state and the desired state and is calculated using the L2-norm.

    In addition to this bound, one of the main contributions of Yi and Crosson was the observation that the leading error term resulting from a first-order Suzuki-Trotter decomposition vanishes for Hamiltonians with certain underlying structures.
    The electronic structure Hamiltonians that we examine in this work have this certain structure.
    As Yi and Crosson note, this result implies that the phase error (and by extension the Trotter error) for a first-order Suzuki-Trotter decomposition has the same asymptotic scaling with respect to the evolution time as a second-order Suzuki-Trotter decomposition when the Hamiltonian has the assumed structure.

    \subsection{Approximating Trotter Error}

    Although the relationship between the phase error and the Trotter error was established by Yi and Crosson, here we will explicitly show how the phase error provides an approximation to the Trotter error (Eq. \ref{methods-eq:trotter-error}) when the state in question is an eigenstate.
    We do not make any assumptions about the order of the Suzuki-Trotter decomposition nor the structure of the Hamiltonians we analyze.

    We begin by writing the target eigenstate ($\ket{\lambda_k}$) in terms of the eigenbasis of the effective Hamiltonian:
    \begin{equation}
        \label{appendix-eq:effective-eigenbasis-expansion}
        \ket{\lambda_k} = \alpha_k \ket{\lambda_k^\prime} + \sum_{j \neq k} \alpha_j \ket{\lambda_j^\prime}
    \end{equation}
    where $\ket{\lambda_k^\prime}$ is the effective eigenstate with the largest squared overlap with the target eigenstate and the eigenstates $\ket{\lambda_j^\prime}$ are the remaining effective eigenstates.

    Computing the phase error with respect to the target eigenstate leads to:
    \begin{equation}
    \begin{split}
        \theta_{\lambda_k} &=  \arg \Big[ \bra{\lambda_k} U^{\dagger}(t) U^\prime(t) \ket{\lambda_k} \Big] =  \arg \Big[ e^{i \lambda_k t} \bra{\lambda_k} U^\prime(t) \ket{\lambda_k} \Big] \\
            &=  \arg \Big[ e^{i \lambda_k t} \bra{\lambda_k} \big(\alpha_k U^\prime(t) \ket{\lambda_k^\prime} + \sum_{j \neq k} \alpha_j U^\prime(t) \ket{\lambda_j^\prime}\big) \Big] \\
            &=  \arg \Big[ e^{i \lambda_k t} \big(\alpha_k e^{-i \lambda^\prime_k t} \braket{\lambda_k}{\lambda_k^\prime} + \sum_{j \neq k} \alpha_j e^{-i \lambda^\prime_j t} \braket{\lambda_k}{\lambda_j^\prime}\big) \Big] \\
            &=  \arg \Big[ e^{i \lambda_k t} \big(|\alpha_k|^2 e^{-i \lambda^\prime_k t} + \sum_{j \neq k} |\alpha_j|^2 e^{-i \lambda^\prime_j t} \big) \Big] \\
            &=  \arg \Big[ |\alpha_k|^2 e^{it (\lambda_k - \lambda^\prime_k)} + \sum_{j \neq k} |\alpha_j|^2 e^{it (\lambda_k - \lambda^\prime_j)} \Big] \\
    \end{split}
    \end{equation}

    When $\sum_{i}|\alpha_j|^2 \ll |\alpha_k|^2$ (which will be satisfied for sufficiently small values of $t$), then this leads to an approximation of the Trotter error:
    \begin{equation}
        \label{appendix-eq:phase-approximation}
        t^{-1} \theta_{\lambda_k} \approx t^{-1} \arg \Big[ |\alpha_k|^2 e^{it (\lambda_k - \lambda^\prime_k)} \Big] = \lambda_k - \lambda^\prime_k
    \end{equation}
    A similar approximation is obtained when computing the phase error with respect to the effective eigenstate.

    It should be noted that Eq. \ref{appendix-eq:phase-approximation} does not lead to an upper-bound of $\epsilon^\text{TS}_k$.
    However, it will give a close approximation to $\epsilon^\text{TS}_k$ in the regime where the fidelity error and the phase error are relatively small (when $t$ is small).

    Eq. \ref{eq:phase-error-approximates-trotter-error} only uses the information about the magnitude of the phase error.
    However, Eq. \ref{appendix-eq:phase-approximation} also computes the sign of $\epsilon^\text{TS}_k$.
    This additional information about the direction of the offset of the eigenvalue could be used when post-processing estimates of the eigenvalues from QPE.
    We do not focus on improving the error analysis of QPE using this sign information, though we note that this may be an interesting direction for future work.

    \subsection{Non-Eigenstates}

    Here, we discuss the use of the phase error when computed with respect to an arbitrary quantum state that has some non-zero overlap with the target eigenstate.

    Let an arbitrary quantum state be defined by:
    \begin{equation}
        \label{eq:generating-state-with-desired-fidelity}
        \ket{\psi} = \sqrt{f} \ket{\lambda_k} + \sum_{j \neq k} \sqrt{\alpha_j} \ket{\lambda_j}
    \end{equation}
    where $f$ is the squared overlap with the target eigenstate ($\ket{\lambda_k}$), $\ket{\lambda_j}$ are the remaining eigenstates, and $\sqrt{\alpha_j}$ are complex-valued coefficients.
    The values of the coefficients of the orthogonal eigenstates are normalized such that: $\sum_{j \neq k} |\alpha_j|^2 = 1 - f$.

    Computing the phase error with respect to the state $\ket{\psi}$ gives:
    \begin{equation}
        \begin{split}
        \theta_\psi &= \arg \Big( \big[\sqrt{f} \bra{\lambda_k} + \sum_{j \neq k} \sqrt{\alpha_j^*} \bra{\lambda_j} \big] U^\dagger(t) U^\prime(t) \big[\sqrt{f} \ket{\lambda_k} + \sum_{j \neq k} \sqrt{\alpha_j} \ket{\lambda_j} \big] \Big) \\
        &= \arg \Big( f e^{i \lambda_k t} \bra{\lambda_k} U^\prime(t) \ket{\lambda_k} + \sum_{j \neq k} |\alpha_j| e^{i \lambda_j t} \bra{\lambda_j} U^\prime(t) \ket{\lambda_j}  + \text{cross-terms} \Big)
        \end{split}
    \end{equation} 
    If we assume that the evolution time is small enough that the eigenbasis of $U(t)$ and $U^\prime(t)$ are nearly identical (the corresponding eigenstates have fidelities close to $1$), then we can make the following approximation where all cross-terms vanish:
    \begin{equation}
        \label{appendix-eq:phase-error-arbitrary-state}
            \theta_\psi \approx \arg \Big( f e^{i (\lambda_k - \lambda^\prime_k) t} + \sum_{j \neq k} |\alpha_j| e^{i (\lambda_j - \lambda^\prime_j) t} \Big)
    \end{equation}  

    Following Eq. \ref{appendix-eq:phase-error-arbitrary-state}, the phase error of an arbitrary quantum state includes contributions from all eigenstates that have non-negligible overlap with the input state.
    Since the phase errors on individual eigenstates can be positive or negative, these phases can then constructively or destructively interfere, meaning the total phase error can be zero.  
    As the fidelity between the input state and the target eigenstate increases, the phase error gives a better approximation to the Trotter error.

    In practice, one may not have access to the target eigenstate, but instead some state that has a nonzero-overlap with the target eigenstate.
    One solution to address this constraint would be to measure the phase error for several states with increasing overlap with the target eigenstate, such as matrix product states with larger bond-dimension, and then extrapolate with respect to the overlap.

    \section{Models}
    \label{appendix:models}

    \subsection{Random Pauli Hamiltonians}

    The randomly sampled Pauli Hamiltonians used in this work are given by a linear combination of randomly generated Pauli operators.
    To generate a Hamiltonian with $L$ terms acting on $N$ qubits, we first build the set of all $4^N$ possible tensor products of the single-qubit operators, $X$, $Y$, $Z$, or $I$:
    \begin{equation}
        X = \begin{pmatrix}
            0 & 1 \\
            1 & 0
            \end{pmatrix}
        \hspace{3em}
        Y = \begin{pmatrix}
            0 & -i \\
            i & 0
        \end{pmatrix}
        \hspace{3em}
        Z = \begin{pmatrix}
            1 & 0 \\
            0 & -1
            \end{pmatrix}
        \hspace{3em}
        I = \begin{pmatrix}
            1 & 0 \\
            0 & 1
            \end{pmatrix}
    \end{equation}

    We then sample $L$ unique Pauli operators from this set and build the Hamiltonian as the linear combination of these operators, each with a coefficient of $1$. 
    Unless otherwise stated, we set $L = N^2$.

    \subsection{Chemistry (Electronic Structure)}

    We investigate the electronic structure Hamiltonians of several small molecules in second-quantization.
    The creation and annihilation operators are mapped to Pauli operators by the Jordan-Wigner \cite{jordan1928paulische} transformation.
    We consider both tapered and untapered Hamiltonians generated by Qubit tapering \cite{bravyi2017tapering}.
    The data files for the Hamiltonians can be found within the open-source software package: Symmer \cite{kirby2021contextual,weaving2023stabilizer,ralli2023unitary,weaving2023contextual}.
    We include the molecules, basis sets, multiplicities, numbers of qubits, and numbers of terms for each Hamiltonian in Table \ref{extended-data-table:chemistry_hamiltonians}.

    One chemistry Hamiltonian was excluded from our analysis: Oxygen in the sto-3g basis with triplet multiplicity under the Jordan-Wigner transformation.
    This model was excluded because the Trotter error on the ground-state ($\epsilon^\text{TS}_0$) was computed to be less than $1e^{-12}$.
    This numerical value is small enough for classical floating point operations to impact our numerical results.

    \begin{center}
        \begin{table}[t]
            \vspace*{\fill}
            \begin{tabular}{|| ccccccc ||}
                \hline
                Molecule & Basis & Multiplicity & $N$ & $L$ & $N^*$ & $L^*$\\ [0.5ex] 
                \hline\hline
                H$_3^+$ & sto-3g & singlet & 6 & 52 & 3 & 34 \\ [0.5ex] \hline
                H$_4$ & sto-3g & singlet & 8 & 105 & 4 & 62 \\ [0.5ex] \hline
                H$_2$ & 3-21g & singlet & 8 & 185 & 5 & 122 \\ [0.5ex] \hline
                H$_2$ & 6-31g & singlet & 8 & 185 & 5 & 122 \\ [0.5ex] \hline
                HeH$^+$ & 3-21g & singlet & 8 & 361 & 6 & 319 \\ [0.5ex] \hline
                Li & sto-3g & doublet & 10 & 156 & 5 & 98 \\ [0.5ex] \hline
                Be & sto-3g & singlet & 10 & 156 & 5 & 102 \\ [0.5ex] \hline
                B$^+$ & sto-3g & singlet & 10 & 156 & 5 & 102 \\ [0.5ex] \hline
                B & sto-3g & doublet & 10 & 156 & 5 & 95 \\ [0.5ex] \hline
                C & sto-3g & triplet & 10 & 156 & 5 & 102 \\ [0.5ex] \hline
                N & sto-3g & quartet & 10 & 156 & 5 & 95\\ \hline
                O & sto-3g & triplet & - & - & 5 & 102 \\ [0.5ex] \hline
                LiH & sto-3g & singlet & 12 & 631 & 8 & 558 \\ [0.5ex] \hline
                BeH$^+$ & sto-3g & singlet & 12 & 631 & 8 & 558 \\ [0.5ex] \hline
                BH & sto-3g & singlet & 12 & 631 & 8 & 558 \\ [0.5ex] \hline
                CH$^+$ & sto-3g & singlet & 12 & 631 & 8 & 558 \\ [0.5ex] \hline
                NH & sto-3g & singlet & 12 & 631 & 8 & 558 \\ [0.5ex] \hline
                OH$^-$ & sto-3g & singlet & 12 & 631 & 8 & 558 \\ [0.5ex] \hline
                HF & sto-3g & singlet & 12 & 631 & 8 & 558 \\ [0.5ex] \hline
                NeH$^+$ & sto-3g & singlet & 12 & 631 & 8 & 558 \\ [0.5ex] \hline
                H$_3^+$ & 3-21g & singlet & 12 & 799 & 9 & 790 \\ [0.5ex] \hline
                H$_2$ & 6-311g & singlet & 12 & 919 & 9 & 919 \\ [0.5ex] \hline
                H$_6$ & sto-3g & singlet & 12 & 919 & 9 & 919 \\ [0.5ex] \hline
                HeH$^+$ & 6-311g & singlet & 12 & 1819 & 10 & 1819 \\ [0.5ex] \hline
                BeH$_2$ & sto-3g & singlet & 14 & 666 & 9 & 596 \\ [0.5ex] \hline
                BH$_2^+$ & sto-3g & singlet & 14 & 1086 & 10 & 1035 \\ [0.5ex] \hline
                CH$_2$ & sto-3g & triplet & 14 & 1086 & 10 & 1035 \\ [0.5ex] \hline
                NH$_2^-$ & sto-3g & singlet & 14 & 1086 & 10 & 1035 \\ [0.5ex] \hline
                H$_2$O & sto-3g & singlet & 14 & 1086 & 10 & 1035 \\ [0.5ex] \hline
            \end{tabular}
            \caption{
                \textbf{Chemistry Hamiltonians}
                Relevant information regarding the electronic structure Hamiltonians is given.
                In order from right to left: the chemical formula of the molecule, the molecular orbital basis, the multiplicity of the molecule, the number of qubits for the \textit{untapered} Hamiltonian ($N$), the number of Pauli operators in the \textit{untapered} Hamiltonian ($L$), the number of qubits for the \textit{tapered} Hamiltonian ($N^*$), and the number of Pauli operators in the \textit{tapered} Hamiltonian ($L^*$).
            }
            \label{extended-data-table:chemistry_hamiltonians}
            \vspace*{\fill}
        \end{table}
    \end{center}

    \section{Trotter Error Scaling}
    \label{appendix:trotter_error_constant}

    \begin{center}
        \begin{figure}[H]
            \centering
            \includegraphics[width=0.75\linewidth]{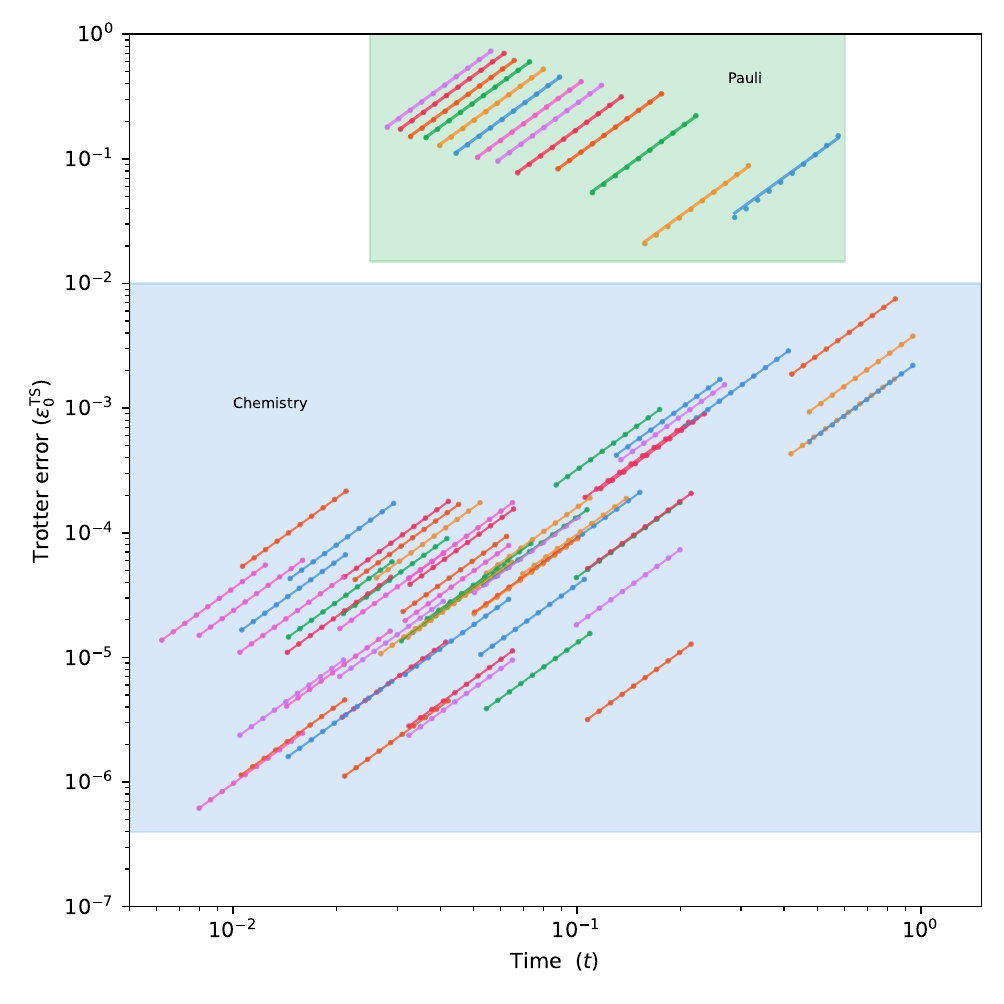}
            \caption{\textbf{Trotter Error Scaling}
            The Trotter error is shown as a function of evolution time ($t$).
            The random Pauli Hamiltonians are located in the shaded green region and the chemistry systems are located in the shaded blue region.
            The Trotter error on the ground state ($\epsilon^\text{TS}_0$) for a second-order ($p=2$) Suzuki-Trotter decomposition is depicted by the circles.
            A quadratic fit for each model is depicted by the solid lines and the minimum $r^2$ value among all models is 0.9996 (0.9927) for the chemistry systems (random Pauli Hamiltonians).
            }
            \label{appendix-fig:trotter_error_scaling}
        \end{figure}
    \end{center}

    In this work, we assume that the Trotter error scales as $W_k t^p$ where $W_k$ is a constant with respect to $t$ for a $p^\text{th}$-order Suzuki-Trotter decomposition.
    This assumption can unfounded in certain cases, particularly for state-dependent error metrics or for small values of $t$ \cite{burgarth2024strong}.
    Since we quantify the Trotter error with respect to a particular eigenstate, $\epsilon^\text{TS}_k$, we numerically validate that this assumption holds in the ranges of $t$ that we consider.

    In Figure \ref{appendix-fig:trotter_error_scaling}, we analyze the scaling of the Trotter error on the ground-state ($\epsilon^\text{TS}_0$) with respect to $t$ for second-order Suzuki-Trotter decompositions ($p=2$).
    $\epsilon^\text{TS}_0$ is computed for $10$ discrete values of $t$, which are evenly spaced on a logarithmic scale within the range $\pi /||4H|| \leq t \leq \pi /||2H||$.
    A quadratic function, $\epsilon^\text{TS}_k \approx W_k t^2$, is fit to these data points to determine the value of $W_k$ associated with each model.
    For all of the models used in this work, the assumption that $\epsilon^\text{TS}_0$ scales as $W_k t^2$ holds as indicated by the quality of the curve fits.
    The minimum $r^2$ values for the random Pauli Hamiltonians and the chemistry systems are $0.9927$ and $0.9996$, respectively.

    It is worth clarifying that this assumption is numerically validated only for second-order Suzuki-Trotter decompositions and within the range of evolution times examined.
    To validate this assumption in practice for large systems, one could measure the approximate phase error for several values of $t$ using the particular approximate time evolution operator being examined.
    These data points could then be fit to validate that the Trotter error scales as $\mathcal{O}(W_k t^p)$ and compute an approximation to $W_k$.

    It is possible that the resource estimation routine used for the algorithm being analyzed would suggest using an evolution time outside of the examined range.
    In this case, the approximate phase error can be measured for the suggested evolution time to determine if the Trotter error constant ($W_k$) is consistent for this value of $t$. 

    \section{Finite Sampling}
    \label{appendix:finite-sampling}

    \begin{center}
        \begin{figure}[H]
            \centering
            \includegraphics[width=\linewidth]{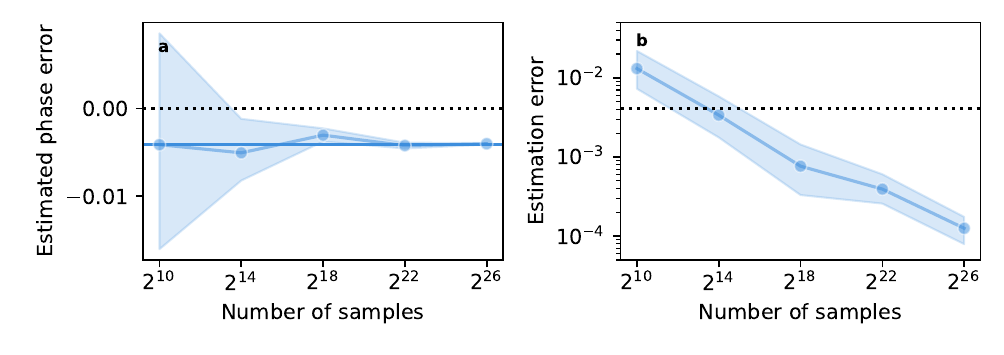}
            \caption{\textbf{Finite Sampling}
                \textbf{a,}
                The uncertainty associated with estimating $\tilde{\theta}_\psi$ is shown as a function of the number of samples.
                The estimate of the approximate phase error is depicted by the blue circles.
                The shaded blue regions depict the confidence interval on the estimate with a confidence level of $95\%$.
                The expected value of the approximate phase error using infinitely many samples is depicted by the solid blue line. 
                The dotted black line depicts the value zero. 
                Results are shown for the ground state of HeH$^+$ in the 3-21g basis ($N=8$).
                The target approximation is a first-order Suzuki-Trotter decomposition ($p=1$) with $t = \pi / ||H||$ and the reference approximation is a fourth-order Suzuki-Trotter decomposition ($p^\prime=4$).
                \textbf{b,} 
                The magnitude of the distance between the estimated value and the expected value of $\tilde{\theta}_\psi$ is shown as a function of the number of samples. 
                The shaded blue regions depict the confidence interval on the estimate with a confidence level of $95\%$. 
                The dotted black line depicts the zero-value estimate.
                Results are shown for the ground state of HeH$^+$ in the 3-21g basis ($N=8$).
                The target approximation is a first-order Suzuki-Trotter decomposition ($p=1$) with $t = \pi / ||H||$ and the reference approximation is a fourth-order Suzuki-Trotter decomposition ($p^\prime=4$).
            }
            \label{appendix-fig:finite-sampling}
        \end{figure}
    \end{center}

    Here, we briefly analyze how many times each circuit in Figure \ref{fig:trotter-had-test} needs to be repeated to produce an estimate of $\tilde{\theta}_\psi$ with a reasonably small uncertainty.
    We refer to the number of independent circuit repetitions as ``shots" or ``samples".

    Using the Hadamard test shown in Figure \ref{fig:trotter-had-test}, the real and imaginary parts of the amplitude encoding the phase error are computed independently:
    \begin{equation}
        \begin{split}
            x \equiv \Re{\bra{\psi}U_\text{ref}^\dagger(t)U_{H^\prime}(t)\ket{\psi}} \\
            y \equiv \Im{\bra{\psi}U_\text{ref}^\dagger(t)U_{H^\prime}(t)\ket{\psi}}
        \end{split}
    \end{equation}
    Given $x$ and $y$, $\tilde{\theta}_\psi$ is computed by:
    \begin{equation}
        \tilde{\theta}_\psi = \tan^{-1}(y/x)
    \end{equation}

    With a finite number of samples, the estimated values of the real ($x_\text{est}$) and imaginary ($y_\text{est}$) parts of the amplitude will have associated statistical uncertainties of $\delta_x$ and $\delta_y$, respectively.
    When an estimate of $\tilde{\theta}_\psi$ is produced from these estimated values, the estimate of $\tilde{\theta}_\psi$ will also have an associated statistical uncertainty ($\delta_\theta$), which will decrease with the number of samples.
    Therefore, the desired uncertainty on the estimated value of $\tilde{\theta}_\psi$ dictates the number of samples that are required.

    We can compute the uncertainty on $\tilde{\theta}_\psi$ ($\delta_\theta$) given the statistical uncertainties of $x_\text{est}$ ($\delta_x$) and $y_\text{est}$ ($\delta_y$):
    \begin{equation}
        \delta_\theta = \sqrt{(\frac{\partial \tilde{\theta}_\psi}{\partial x}\delta_x)^2 + (\frac{\partial \tilde{\theta}_\psi}{\partial y}\delta_y)^2} = (x^2 + y^2)^{-1} \sqrt{y^2 \delta_x^2 + x^2 \delta_y^2}
    \end{equation}
    If we assume that the same number of samples are used to produce $x_\text{est}$ and $y_\text{est}$, then the statistical uncertainties will be equal: $\delta_x = \delta_y = \delta$.
    This leads to:
    \begin{equation}
        \begin{split}
            \delta_\theta = (x^2 + y^2)^{-1/2} \delta
        \end{split}
    \end{equation}
    Therefore, using $\mathcal{O}(1/\delta^2)$ samples to estimate both $x$ and $y$ independently will produce an estimate of $\tilde{\theta}_\psi$ with a statistical uncertainty of $\mathcal{O}(\delta)$.

    Numerically, we show the estimated value of $\tilde{\theta}_\psi$ as a function of the number of samples in Figure \ref{appendix-fig:finite-sampling} for HeH$^+$ in the 3-21G basis ($N=8$).
    From this data, we see that estimates of the approximate phase error that result in reasonably high confidence can be achieved with fewer than $10^6$ samples for this model.
    We do not expect that the required number of samples will increase with system size and the desired precision will be highly dependent upon the context in which the simulation error is being used.

    \section{Accuracy of Phase Error Approximation}
    \label{appendix:phase_approximation}

    \begin{center}
        \begin{figure}[H]
            \centering
            \includegraphics[width=\linewidth]{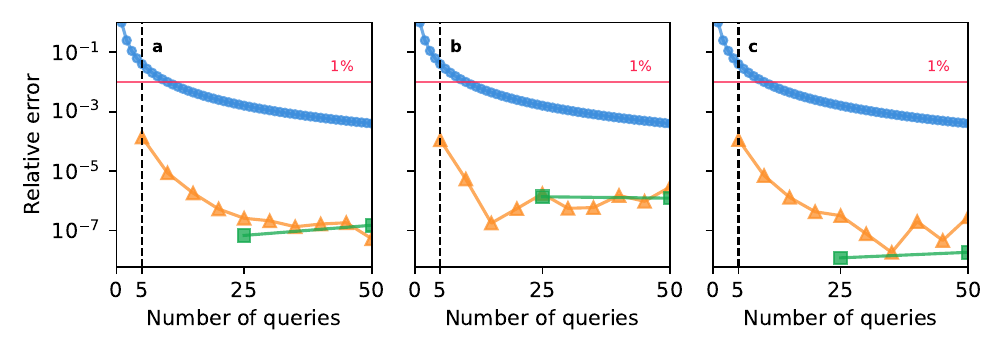}
            \caption{\textbf{Approximating Phase Error with Multiple Trotter Steps}
            The relative error on the approximate phase error ($\tilde{\theta}_\psi$) is shown as a function of the number of queries that are used to generate the reference approximation for three chemistry systems: \textbf{a,} C ($N = 10$), \textbf{b,} HF ($N = 12$), and \textbf{(c)} NH$_2^-$ ($N = 14$).
            The number of queries is the number of individual calls to a second-order Suzuki-Trotter decomposition.
            The target approximation is a second-order ($p=2$) Suzuki-Trotter decomposition with $t = \pi / ||4H||$.
            Reference approximations with various numbers of Trotter steps ($S$) are depicted by the blue circles ($p^\prime = 2$), orange triangles ($p^\prime = 4$), and green squares ($p^\prime = 6$).
            The red line depicts a relative error of $1\%$.
            The dashed black line depicts the fewest number of queries required to generate a reference approximation with less than $1\%$ relative error.
            }
            \label{appendix-fig:phase-approximation}
        \end{figure}
    \end{center}

    \begin{center}
        \begin{table}[t]
            \begin{tabular}{|| ccccc ||}
                \hline
                Molecule & Basis & Multiplicity & Relative error & Relative error$^*$ \\ [0.5ex] 
                \hline\hline
                H$_3^+$ & 3-21g & singlet & 2.87e-05 & 0.00014 \\ [0.5ex] \hline
                H$_4$ & sto-3g & singlet & 0.000292 & 0.000306 \\ [0.5ex] \hline
                H$_2$ & 3-21g & singlet & 0.000106 & 0.000682 \\ [0.5ex] \hline
                H$_2$ & 6-31g & singlet & 0.000113 & 0.000741 \\ [0.5ex] \hline
                HeH$^+$ & 3-21g & singlet & 0.000484 & 0.00166 \\ [0.5ex] \hline
                Li & sto-3g & doublet & 0.000541 & 0.000464 \\ [0.5ex] \hline
                Be & sto-3g & singlet & 0.000188 & 0.000214 \\ [0.5ex] \hline
                B$^+$ & sto-3g & singlet & 0.000177 & 0.000391 \\ [0.5ex] \hline
                B & sto-3g & doublet & 0.000171 & 0.000385 \\ [0.5ex] \hline
                C & sto-3g & triplet & 0.000139 & 0.000461 \\ [0.5ex] \hline
                N & sto-3g & quartet & 1.19e-05 & 0.000513 \\ [0.5ex] \hline
                O & sto-3g & triplet & - & 0.000454 \\ [0.5ex] \hline
                LiH & sto-3g & singlet & 0.000168 & 0.000399 \\ [0.5ex] \hline
                BeH$^+$ & sto-3g & singlet & 0.000185 & 0.00048 \\ [0.5ex] \hline
                BH & sto-3g & singlet & 0.000187 & 0.000313 \\ [0.5ex] \hline
                CH$^+$ & sto-3g & singlet & 0.00014 & 0.000245 \\ [0.5ex] \hline
                NH & sto-3g & singlet & 0.000127 & 0.000353 \\ [0.5ex] \hline
                OH$^-$ & sto-3g & singlet & 9.77e-05 & 0.000522 \\ [0.5ex] \hline
                HF & sto-3g & singlet & 0.000112 & 0.000515 \\ [0.5ex] \hline
                NeH$^+$ & sto-3g & singlet & 0.00349 & 0.000496 \\ [0.5ex] \hline
                H$_3^+$ & sto-3g & singlet & 0.000836 & 0.00155 \\ [0.5ex] \hline
                H$_2$ & 6-311g & singlet & 4.13e-05 & 0.000171 \\ [0.5ex] \hline
                H$_6$ & sto-3g & singlet & 9.08e-05 & 0.000754 \\ [0.5ex] \hline
                HeH$^+$ & 6-311g & singlet & 0.000178 & 0.000543 \\ [0.5ex] \hline
                BeH$_2$ & sto-3g & singlet & 0.000128 & 0.000202 \\ [0.5ex] \hline
                BH$_2^+$ & sto-3g & singlet & 0.000185 & 0.000358 \\ [0.5ex] \hline
                CH$_2$ & sto-3g & triplet & 0.000155 & 0.000241 \\ [0.5ex] \hline
                NH$_2^-$ & sto-3g & singlet & 0.000114 & 0.000495 \\ [0.5ex] \hline
                H$_2$O & sto-3g & singlet & 0.000128 & 0.000537 \\ [0.5ex] \hline
            \end{tabular}
            \caption{
                \textbf{Relative Error (Chemistry)}
                The relative error on the approximate phase error ($\tilde{\theta}_\psi$) is shown for the chemistry systems examined in this work.
                The target approximation is a second-order Suzuki-Trotter decomposition ($p = 2$) and the reference approximation is a fourth-order Suzuki-Trotter decomposition ($p^\prime = 4$) with $t = \pi / ||4H||$.
                ``Relative error" and ``Relative error$^*$" correspond to the untapered and tapered systems, respectively.
            }
            \label{extended-data-table:approximate-phase-error-error-chemistry}
        \end{table}
    \end{center}

    \begin{center}
        \begin{table}[t]
            \begin{tabular}{|| ccc ||}
                \hline
                $N$ & $L$ & Relative error \\ [0.5ex] 
                \hline\hline
                2 & 4 & 0.0077 \\ \hline
                3 & 9 & 0.00404 \\ \hline
                4 & 16 & 0.0036 \\ \hline
                5 & 25 & 0.00321 \\ \hline
                6 & 36 & 0.00295 \\ \hline
                7 & 49 & 0.00376 \\ \hline
                8 & 64 & 0.00317 \\ \hline
                9 & 81 & 0.00354 \\ \hline
                10 & 100 & 0.00382 \\ \hline
                11 & 121 & 0.00389 \\ \hline
                12 & 144 & 0.0036 \\ \hline
                13 & 169 & 0.00372 \\ \hline
                14 & 196 & 0.0036 \\ \hline
            \end{tabular}
            \caption{
                \textbf{Relative Error (random Pauli)}
                The relative error on the approximate phase error ($\tilde{\theta}_\psi$) is shown for the random Pauli Hamiltonians examined in this work.
                The target approximation is a second-order Suzuki-Trotter decomposition ($p = 2$) and the reference approximation is a fourth-order Suzuki-Trotter decomposition ($p^\prime = 4$) with $t = \pi / ||4H||$.
            }
            \label{extended-data-table:approximate-phase-error-error-random}
        \end{table}
    \end{center}

    Here, we expand on our analysis concerning the error associated when approximating the phase error via Eq. \ref{eq:approximate-phase-error}.
    Let the reference approximation be generated by a ${p^\prime}^\text{th}$-order Suzuki-Trotter decomposition using $S$ Trotter steps: $U_\text{ref}^\dagger = [U_{p^\prime}^\prime(t/S)]^S$.
    As noted, we can increase the accuracy of $\tilde{\theta}_\psi$ by increasing the accuracy of the reference approximation, at the expense of requiring more operations to implement the reference approximation.

    We would like to determine the lowest-cost reference approximation that results in an approximate phase error that is within a certain, tolerable amount of error.
    Higher-order Suzuki-Trotter decompositions are constructed using recursive calls to lower-order decompositions, with the base case being the second-order decomposition (Eq. \ref{appendix-eq:higher-order-trotter-decompositions}).
    This allows us to quantify the cost of each reference approximation using the number of calls to the second-order Suzuki-Trotter decomposition.

    In Figure \ref{appendix-fig:phase-approximation}, we show the relative error of $\tilde{\theta}_{\lambda_0}$ for different values of $p^\prime$ and $S$ as a function of the total number of queries to the second-order Suzuki-Trotter decomposition when the target approximation is a second-order Suzuki-Trotter decomposition ($p = 2$).
    We show $3$ representative models with differing system sizes, though we note that all of the models examined in this work showed similar trends.
    We chose $1\%$ relative error as a ``tolerable" amount of error for the approximation of the phase error.

    Numerically, we determined that $p^\prime = 4$ and $S = 1$ lead to the lowest cost reference approximation that resulted in in less than $1\%$ relative error for every model that we considered.
    The relative errors associated with $p^\prime = 4$ and $S = 1$ for each model are given in Tables \ref{extended-data-table:approximate-phase-error-error-chemistry} and \ref{extended-data-table:approximate-phase-error-error-random}.
    This data suggests that the order of the reference approximation needed to adequately approximate $\theta_\psi$ is independent of system size, and that the number of resources required to compute $\tilde{\theta}_\psi$ will scale only with the number of terms in the Hamiltonian ($L$).

    As noted in Section \ref{sec:algorithm}, the reference approximation incurs an additional error term, which can lead to either an underestimate or overestimate of $\theta_\psi$.
    In subfigure \ref{appendix-fig:extrapolating_approximations}a, we show $\tilde{\theta}_\psi$ as a function of the order of the reference approximation ($p^\prime$) for HeH$^+$ in the 3-21G basis ($N=8$).
    In this case, the magnitude of $\tilde{\theta}_\psi$ underestimates the magnitude of $\theta_\psi$ for small values of $p^\prime$.

    \section{Quantum Computing Fidelity Error}
    \label{appendix:fidelity-error}

    In this work, we focus on computing $\theta_\psi$ to produce numerical resource estimates for Trotter-based QPE.
    Resource estimates for other quantum simulation algorithms - such as those for analyzing the dynamics of quantum systems - may additionally depend on the fidelity error ($f_\psi$).
    Here, we show how to approximate $f_\psi$ using the same measurement results used to compute $\theta_\psi$.

    The Hadamard test outlined in Figure \ref{fig:trotter-had-test} depicts measuring the real and imaginary parts of the amplitude $\bra{\psi}U_\text{ref}^\dagger(t) U_{H^\prime}(t)\ket{\psi}$.
    The estimates of the real ($x$) and imaginary ($y$) parts of this amplitude are converted into an estimate of the phase of the amplitude: 
    \begin{equation}
        \label{appendix-eq:amplitude-to-phase}
        \arg(x + iy) \approx \arg(\sqrt{1-f_\psi} e^{i\theta_\psi}) = \theta_\psi
    \end{equation}
    The above amplitude information also encodes the fidelity error ($f_\psi$):
    \begin{equation}
        1 - |x|^2 - |y|^2 \approx f_\psi 
    \end{equation}
    Therefore, the estimates of $x$ and $y$ can be used to produce an approximation of $f_\psi$, without requiring any additional quantum resources.

    \begin{center}
        \begin{figure}[H]
            \centering
            \includegraphics[width=\linewidth]{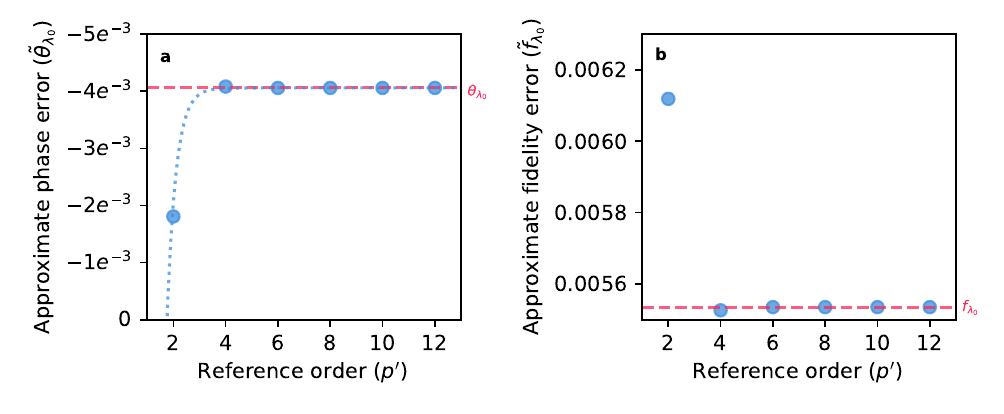}
            \caption{
                \textbf{Approximating Simulation Errors}
                \textbf{a,} Approximations of the phase error ($\tilde{\theta}_\psi$) are shown as a function of the order of the Suzuki-Trotter decomposition used for the reference approximation ($p^\prime$).
                Results are shown for the ground state of HeH$^+$ in the 3-21g basis ($N = 8$).
                The target approximation is a first-order Suzuki-Trotter decomposition ($p=1$) with $t = \pi / ||H||$.
                The dashed red line depicts the phase error on the ground state ($\theta_{\lambda_0}$).
                The dotted blue line depicts a curve fit where the error of the approximation decays exponentially.
                \textbf{b,} Approximations of the fidelity error ($\tilde{f}_\psi$) are shown as a function of the order of the Suzuki-Trotter decomposition used for the reference approximation ($p^\prime$).
                Results are shown for the ground state of HeH$^+$ in the 3-21g basis ($N = 8$).
                The target approximation is a first-order Suzuki-Trotter decomposition ($p=1$) with $t = \pi / ||H||$.
                The dashed red line depicts the fidelity error on the ground state ($f_{\lambda_0}$).
            }
            \label{appendix-fig:extrapolating_approximations}
        \end{figure}
    \end{center}

    In subfigure \ref{appendix-fig:extrapolating_approximations}b, we show the accuracy of the approximation to $f_\psi$ using the same system and reference approximation for the data displayed in subfigure \ref{appendix-fig:extrapolating_approximations}a.
    The error on the approximation of $f_\psi$ decays rapidly as a function of $p^\prime$.
    The relative error on the approximation of $f_\psi$ when $p^\prime = 4$ is approximately $0.16\%$ and is approximately $0.00037\%$ when $p^\prime = 6$.

    Since the fidelity error can be computed using the same measurement results used to compute $\theta_\psi$, resource estimates that rely on $f_\psi$ can be obtained without using any additional quantum computation.
    Alternatively, if one only wishes to compute either $\theta_\psi$ or $f_\psi$, it may be possible to reduce the required quantum resources since additional information is being obtained.
    This would be achieved by altering the amplitude estimation algorithm such that only the information corresponding to the phase of the ampltidue ($\theta_\psi$) or the magnitude of the amplitude ($f_\psi$) is extracted.
    We leave such adaption and analysis for future work.

    \section{Extrapolating Along State Preparation}
    \label{appendix:fidelity-extrapolation}

    \begin{center}
        \begin{figure}[H]
            \centering
            \includegraphics[width=\linewidth]{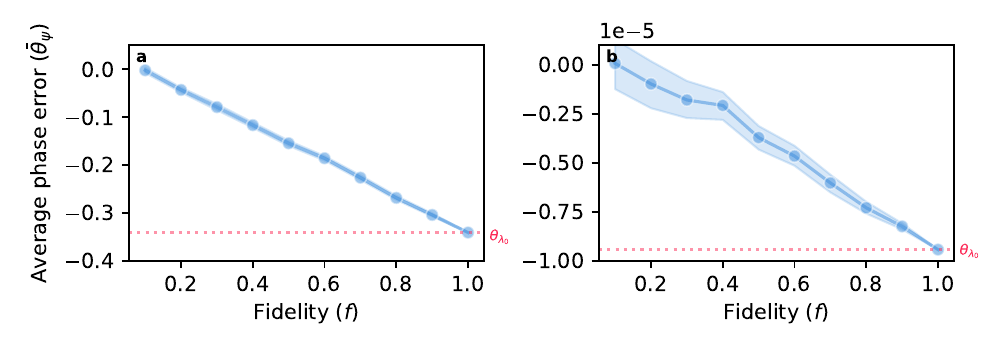}
            \caption{\textbf{Extrapolating Along Fidelity}
            \textbf{a,}
            The average phase error ($\bar{\theta}_\psi$) is shown as a function of fidelity ($f$) with the ground-state.
            Results are shown for a randomly sampled Pauli Hamiltonian with $L=64$ terms and $N=8$ qubits.
            The blue circles depict the average phase error ($\bar{\theta}_\psi$) computed over $100$ randomly generated states with the chosen fidelity.
            The shaded blue regions depict the confidence interval on the average with a confidence level of $95\%$.
            The dashed red line depicts the phase error on the ground state ($\theta_{\lambda_0}$).
            The target approximation is a second-order Suzuki-Trotter decomposition ($p=2$) with $t = \pi/||H||$.
            \textbf{b,} 
            The average phase error ($\bar{\theta}_\psi$) is shown as a function of fidelity ($f$) with the ground-state.
            Results are shown for CH$_2$ in the sto-3g basis with triplet multiplicity ($N = 14$).
            The blue circles depict the average phase error ($\bar{\theta}_\psi$) computed over $100$ randomly generated states with the chosen fidelity.
            The shaded blue regions depict the confidence interval on the average with a confidence level of $95\%$.
            The dashed red line depicts the phase error on the ground state ($\theta_{\lambda_0}$).
            The target approximation is a second-order Suzuki-Trotter decomposition ($p=2$) with $t = \pi/||H||$.
            }
            \label{fig:fidelity_extrapolation}
        \end{figure}
    \end{center}

    In practice, preparing eigenstates on the quantum computer may be too costly for NISQ and MegaQuop devices.
    Here, we propose one strategy for addressing this constraint.

    Eq. \ref{appendix-eq:phase-error-arbitrary-state} implies that the phase error on a quantum state ($\ket{\psi}$) will better approximate the phase error on the target eigenstate ($\ket{\lambda_k}$) as the fidelity ($f$) between $\ket{\psi}$ and $\ket{\lambda_k}$ increases.
    Let us assume that we can prepare several quantum states, each with increasing fidelity with the target eigenstate.
    We can then measure the phase error for each state and then extrapolate to estimate the phase error on the target eigenstate.

    In Figure \ref{fig:fidelity_extrapolation}, we show $\theta_\psi$ as a function of $f$ for one randomly sampled Pauli Hamiltonian and one chemistry system.
    For each value of $f$, we compute the average of $\theta_\psi$ for $100$ randomly generated states.
    For both models, we see that $\theta_\psi$ approaches $\theta_{\lambda_0}$ smoothly as $f$ increases.
    However, we note that statistical error will be particularly influential for cases where the magnitude of the phase error is close to zero.

    The assumption that one can prepare several quantum states with increasing fidelity is not unfounded.
    In the case of chemistry Hamiltonians, one could use classical chemistry algorithms with increasing accuracies to generate classical ansatz states.
    Alternatively, one could run several DMRG several times with increasing maximum bond-dimension.
    The resulting matrix product states can be efficiently prepared on the quantum computer following Berry et al. \cite{google_mps_prep} and extrapolation could be performed with respect to the bond-dimension.
    Another strategy could be to prepare several states via Adiabatic State Preparation \cite{farhi2000quantum, babbush2014adiabatic, albash2018adiabatic} and extrapolate with respect to the quantum resources used for state preparation.

    \section{Estimating Operator Norm}
    \label{appendix:estimating-operator-norm}

    \begin{center}
        \begin{figure}[H]
            \centering
            \includegraphics[width=\linewidth]{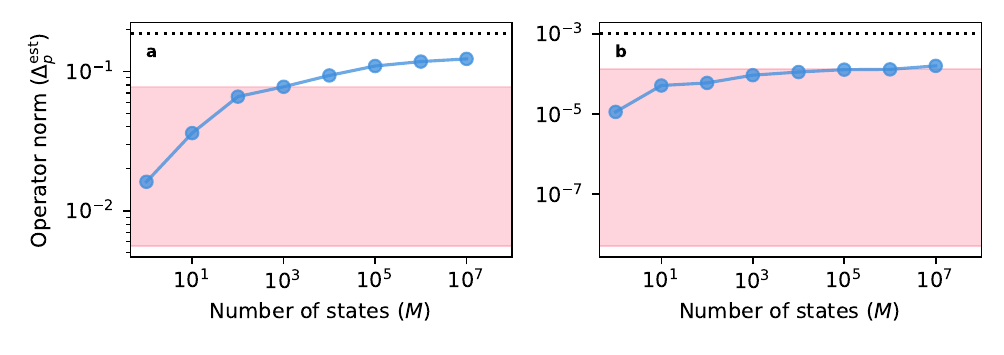}
            \caption{
                \textbf{Estimating Operator Norm}
                \textbf{a,}
                The estimate of the operator norm ($\Delta^\text{est}_p$) is shown as a function of the number of states tested ($M$).
                The blue circles depict the estimated operator norm ($\Delta^\text{est}_p$).
                The dashed black line depicts the expected operator norm ($\Delta_p$).
                The shaded red region depicts the range between the maximum and minimum phase errors on an eigenstate.
                The target approximation is a second-order Suzuki-Trotter decomposition ($p = 2$) with $t = \pi / ||2H||$.
                Results are shown for a random Pauli Hamiltonian with $L=16$ terms and $N = 4$ qubits.
                \textbf{b,}
                The estimate of the operator norm ($\Delta^\text{est}_p$) is shown as a function of the number of states tested ($M$).
                The blue circles depict the estimated operator norm ($\Delta^\text{est}_p$).
                The dashed black line depicts the expected operator norm ($\Delta_p$).
                The shaded red region depicts the range between the maximum and minimum phase errors on an eigenstate.
                The target approximation is a second-order Suzuki-Trotter decomposition ($p = 2$) with $t = \pi / ||2H||$.
                Results are shown for Lithium Hydride in the sto-3g basis with singlet multiplicity and qubit tapering ($N = 8$).
            }
            \label{fig:estimating-operator-norm}
        \end{figure}
    \end{center}

    The operator norm ($\Delta_p$) described by Eq. \ref{appendix-eq:operator-norm} can be used to compute resource estimates that reflect the worst-case over all eigenstates.
    Computing $\Delta_p$ is a non-trivial task as it is an eigenvalue problem.
    Here, we look at estimating $\Delta_p$ using a quantum computer.

    As noted in Appendix \ref{appendix:commutator_bounds}, $\Delta_p$ upper-bounds the maximum separation in the distance between the desired state and the achieved state.
    Consequently, it also upper-bounds the phase error among all quantum states:
    \begin{equation}
        \label{appendix-eq:max-phase-error}
        \Delta_p \equiv || U(t) - U^\prime_p(t) || \geq \max_\psi | \theta_\psi |
    \end{equation} 

    The converse statement - the maximum phase error over all quantum states approximates the operator norm - can be useful if one wishes to determine the worst-case impact of the Trotter error.
    From this perspective, an estimate of $\Delta_p$ can be obtained as the largest phase error of an ensemble of quantum states:
    \begin{equation}
        \label{appendix-eq:op-norm-est}
        \Delta^\text{est}_p = \max_{m \in [0, M)} | \theta_{\psi_m} | 
    \end{equation}
    where $M$ is the number of unique quantum states considered.

    In Figure \ref{fig:estimating-operator-norm}, we show numerically computed values of $\Delta^\text{est}_p$ as a function of $M$.
    We show results for a random Pauli Hamiltonian with $16$ terms acting on $4$ qubits (subfigure a) and Lithium Hydride in the sto-3g basis with singlet multiplicity under the Jordan-Wigner transformation with qubit tapering acting on $8$ qubits (subfigure b). 
    We randomly generate quantum states with no inherent structure and compute $\Delta^\text{est}_p$ following Eq. \ref{appendix-eq:op-norm-est}. 

    For both systems, $\Delta^\text{est}_p$ increases with $M$.
    For the random Pauli Hamiltonian, $\Delta^\text{est}_p$ exceeds the worst-case phase error on an eigenstate after sampling $10^4$ quantum states.
    For Lithium Hydride, $\Delta^\text{est}_p$ requires $10^7$ randomly generated quantum states in order to exceed the worst-case phase error on an eigenstate.

    The total number of eigenstates grows exponentially with system size.
    Therefore, it might be expected that the value of $M$ needed to have high confidence that $\Delta^\text{est}_p$ will upper-bound the worst-case phase error on an eigenstate will grow exponentially with system size.
    However, if $\Delta^\text{est}_p$ is used to approximate the worst-case phase error on an eigenstate - instead of a strict upper-bound - a sufficient value of $M$ may be much smaller.
    For both systems we examine, $M = 10^3$ provides a resaonable approximation of the worst-case phase error on an eigenstate.


\begin{thebibliography}{75}
    \ifx \bisbn   \undefined \def \bisbn  #1{ISBN #1}\fi
    \ifx \binits  \undefined \def \binits#1{#1}\fi
    \ifx \bauthor  \undefined \def \bauthor#1{#1}\fi
    \ifx \batitle  \undefined \def \batitle#1{#1}\fi
    \ifx \bjtitle  \undefined \def \bjtitle#1{#1}\fi
    \ifx \bvolume  \undefined \def \bvolume#1{\textbf{#1}}\fi
    \ifx \byear  \undefined \def \byear#1{#1}\fi
    \ifx \bissue  \undefined \def \bissue#1{#1}\fi
    \ifx \bfpage  \undefined \def \bfpage#1{#1}\fi
    \ifx \blpage  \undefined \def \blpage #1{#1}\fi
    \ifx \burl  \undefined \def \burl#1{\textsf{#1}}\fi
    \ifx \doiurl  \undefined \def \doiurl#1{\url{https://doi.org/#1}}\fi
    \ifx \betal  \undefined \def \betal{\textit{et al.}}\fi
    \ifx \binstitute  \undefined \def \binstitute#1{#1}\fi
    \ifx \binstitutionaled  \undefined \def \binstitutionaled#1{#1}\fi
    \ifx \bctitle  \undefined \def \bctitle#1{#1}\fi
    \ifx \beditor  \undefined \def \beditor#1{#1}\fi
    \ifx \bpublisher  \undefined \def \bpublisher#1{#1}\fi
    \ifx \bbtitle  \undefined \def \bbtitle#1{#1}\fi
    \ifx \bedition  \undefined \def \bedition#1{#1}\fi
    \ifx \bseriesno  \undefined \def \bseriesno#1{#1}\fi
    \ifx \blocation  \undefined \def \blocation#1{#1}\fi
    \ifx \bsertitle  \undefined \def \bsertitle#1{#1}\fi
    \ifx \bsnm \undefined \def \bsnm#1{#1}\fi
    \ifx \bsuffix \undefined \def \bsuffix#1{#1}\fi
    \ifx \bparticle \undefined \def \bparticle#1{#1}\fi
    \ifx \barticle \undefined \def \barticle#1{#1}\fi
    \bibcommenthead
    \ifx \bconfdate \undefined \def \bconfdate #1{#1}\fi
    \ifx \botherref \undefined \def \botherref #1{#1}\fi
    \ifx \url \undefined \def \url#1{\textsf{#1}}\fi
    \ifx \bchapter \undefined \def \bchapter#1{#1}\fi
    \ifx \bbook \undefined \def \bbook#1{#1}\fi
    \ifx \bcomment \undefined \def \bcomment#1{#1}\fi
    \ifx \oauthor \undefined \def \oauthor#1{#1}\fi
    \ifx \citeauthoryear \undefined \def \citeauthoryear#1{#1}\fi
    \ifx \endbibitem  \undefined \def \endbibitem {}\fi
    \ifx \bconflocation  \undefined \def \bconflocation#1{#1}\fi
    \ifx \arxivurl  \undefined \def \arxivurl#1{\textsf{#1}}\fi
    \csname PreBibitemsHook\endcsname
    
    \bibitem[\protect\citeauthoryear{Benioff}{1980}]{benioff1980computer}
    \begin{barticle}
    \bauthor{\bsnm{Benioff}, \binits{P.}}:
    \batitle{The computer as a physical system: A microscopic quantum mechanical hamiltonian model of computers as represented by turing machines}.
    \bjtitle{Journal of statistical physics}
    \bvolume{22},
    \bfpage{563}--\blpage{591}
    (\byear{1980})
    \end{barticle}
    \endbibitem
    
    \bibitem[\protect\citeauthoryear{Feynman}{1982}]{feynman1982simulating}
    \begin{botherref}
    \oauthor{\bsnm{Feynman}, \binits{R.P.}}:
    Simulating physics with computers.
    International Journal of Theoretical Physics
    \textbf{21}(6/7)
    (1982)
    \end{botherref}
    \endbibitem
    
    \bibitem[\protect\citeauthoryear{Lloyd}{1996}]{lloyd1996universal}
    \begin{barticle}
    \bauthor{\bsnm{Lloyd}, \binits{S.}}:
    \batitle{Universal quantum simulators}.
    \bjtitle{Science}
    \bvolume{273}(\bissue{5278}),
    \bfpage{1073}--\blpage{1078}
    (\byear{1996})
    \end{barticle}
    \endbibitem
    
    \bibitem[\protect\citeauthoryear{Georgescu et~al.}{2014}]{qsimreview}
    \begin{barticle}
    \bauthor{\bsnm{Georgescu}, \binits{I.M.}},
    \bauthor{\bsnm{Ashhab}, \binits{S.}},
    \bauthor{\bsnm{Nori}, \binits{F.}}:
    \batitle{Quantum simulation}.
    \bjtitle{Rev. Mod. Phys.}
    \bvolume{86},
    \bfpage{153}--\blpage{185}
    (\byear{2014})
    \doiurl{10.1103/RevModPhys.86.153}
    \end{barticle}
    \endbibitem
    
    \bibitem[\protect\citeauthoryear{Abrams and Lloyd}{1999}]{abrams1999quantum}
    \begin{barticle}
    \bauthor{\bsnm{Abrams}, \binits{D.S.}},
    \bauthor{\bsnm{Lloyd}, \binits{S.}}:
    \batitle{Quantum algorithm providing exponential speed increase for finding eigenvalues and eigenvectors}.
    \bjtitle{Physical Review Letters}
    \bvolume{83}(\bissue{24}),
    \bfpage{5162}
    (\byear{1999})
    \end{barticle}
    \endbibitem
    
    \bibitem[\protect\citeauthoryear{Shor}{1999}]{shor1999polynomial}
    \begin{barticle}
    \bauthor{\bsnm{Shor}, \binits{P.W.}}:
    \batitle{Polynomial-time algorithms for prime factorization and discrete logarithms on a quantum computer}.
    \bjtitle{SIAM review}
    \bvolume{41}(\bissue{2}),
    \bfpage{303}--\blpage{332}
    (\byear{1999})
    \end{barticle}
    \endbibitem
    
    \bibitem[\protect\citeauthoryear{Harrow et~al.}{2009}]{harrow2009quantum}
    \begin{barticle}
    \bauthor{\bsnm{Harrow}, \binits{A.W.}},
    \bauthor{\bsnm{Hassidim}, \binits{A.}},
    \bauthor{\bsnm{Lloyd}, \binits{S.}}:
    \batitle{Quantum algorithm for linear systems of equations}.
    \bjtitle{Physical review letters}
    \bvolume{103}(\bissue{15}),
    \bfpage{150502}
    (\byear{2009})
    \end{barticle}
    \endbibitem
    
    \bibitem[\protect\citeauthoryear{Preskill}{2018}]{preskill2018quantum}
    \begin{barticle}
    \bauthor{\bsnm{Preskill}, \binits{J.}}:
    \batitle{Quantum computing in the nisq era and beyond}.
    \bjtitle{Quantum}
    \bvolume{2},
    \bfpage{79}
    (\byear{2018})
    \end{barticle}
    \endbibitem
    
    \bibitem[\protect\citeauthoryear{Peruzzo et~al.}{2014}]{peruzzo2014variational}
    \begin{barticle}
    \bauthor{\bsnm{Peruzzo}, \binits{A.}},
    \bauthor{\bsnm{McClean}, \binits{J.}},
    \bauthor{\bsnm{Shadbolt}, \binits{P.}},
    \bauthor{\bsnm{Yung}, \binits{M.-H.}},
    \bauthor{\bsnm{Zhou}, \binits{X.-Q.}},
    \bauthor{\bsnm{Love}, \binits{P.J.}},
    \bauthor{\bsnm{Aspuru-Guzik}, \binits{A.}},
    \bauthor{\bsnm{O’brien}, \binits{J.L.}}:
    \batitle{A variational eigenvalue solver on a photonic quantum processor}.
    \bjtitle{Nature communications}
    \bvolume{5}(\bissue{1}),
    \bfpage{4213}
    (\byear{2014})
    \end{barticle}
    \endbibitem
    
    \bibitem[\protect\citeauthoryear{Quantum et~al.}{2020}]{google2020hartree}
    \begin{barticle}
    \bauthor{\bsnm{Quantum}, \binits{G.A.}},
    \bauthor{\bsnm{Collaborators*†}},
    \bauthor{\bsnm{Arute}, \binits{F.}},
    \bauthor{\bsnm{Arya}, \binits{K.}},
    \bauthor{\bsnm{Babbush}, \binits{R.}},
    \bauthor{\bsnm{Bacon}, \binits{D.}},
    \bauthor{\bsnm{Bardin}, \binits{J.C.}},
    \bauthor{\bsnm{Barends}, \binits{R.}},
    \bauthor{\bsnm{Boixo}, \binits{S.}},
    \bauthor{\bsnm{Broughton}, \binits{M.}},
    \bauthor{\bsnm{Buckley}, \binits{B.B.}}, \betal:
    \batitle{Hartree-fock on a superconducting qubit quantum computer}.
    \bjtitle{Science}
    \bvolume{369}(\bissue{6507}),
    \bfpage{1084}--\blpage{1089}
    (\byear{2020})
    \end{barticle}
    \endbibitem
    
    \bibitem[\protect\citeauthoryear{AI}{2023}]{google2023suppressing}
    \begin{barticle}
    \bauthor{\bsnm{AI}, \binits{G.Q.}}:
    \batitle{Suppressing quantum errors by scaling a surface code logical qubit}.
    \bjtitle{Nature}
    \bvolume{614}(\bissue{7949}),
    \bfpage{676}--\blpage{681}
    (\byear{2023})
    \end{barticle}
    \endbibitem
    
    \bibitem[\protect\citeauthoryear{Bluvstein et~al.}{2024}]{bluvstein2024logical}
    \begin{barticle}
    \bauthor{\bsnm{Bluvstein}, \binits{D.}},
    \bauthor{\bsnm{Evered}, \binits{S.J.}},
    \bauthor{\bsnm{Geim}, \binits{A.A.}},
    \bauthor{\bsnm{Li}, \binits{S.H.}},
    \bauthor{\bsnm{Zhou}, \binits{H.}},
    \bauthor{\bsnm{Manovitz}, \binits{T.}},
    \bauthor{\bsnm{Ebadi}, \binits{S.}},
    \bauthor{\bsnm{Cain}, \binits{M.}},
    \bauthor{\bsnm{Kalinowski}, \binits{M.}},
    \bauthor{\bsnm{Hangleiter}, \binits{D.}}, \betal:
    \batitle{Logical quantum processor based on reconfigurable atom arrays}.
    \bjtitle{Nature}
    \bvolume{626}(\bissue{7997}),
    \bfpage{58}--\blpage{65}
    (\byear{2024})
    \end{barticle}
    \endbibitem
    
    \bibitem[\protect\citeauthoryear{Rodriguez et~al.}{2024}]{rodriguez2024experimental}
    \begin{botherref}
    \oauthor{\bsnm{Rodriguez}, \binits{P.S.}},
    \oauthor{\bsnm{Robinson}, \binits{J.M.}},
    \oauthor{\bsnm{Jepsen}, \binits{P.N.}},
    \oauthor{\bsnm{He}, \binits{Z.}},
    \oauthor{\bsnm{Duckering}, \binits{C.}},
    \oauthor{\bsnm{Zhao}, \binits{C.}},
    \oauthor{\bsnm{Wu}, \binits{K.-H.}},
    \oauthor{\bsnm{Campo}, \binits{J.}},
    \oauthor{\bsnm{Bagnall}, \binits{K.}},
    \oauthor{\bsnm{Kwon}, \binits{M.}}, et al.:
    Experimental demonstration of logical magic state distillation.
    arXiv preprint arXiv:2412.15165
    (2024)
    \end{botherref}
    \endbibitem
    
    \bibitem[\protect\citeauthoryear{Lacroix et~al.}{2025}]{lacroix2024scaling}
    \begin{barticle}
    \bauthor{\bsnm{Lacroix}, \binits{N.}},
    \bauthor{\bsnm{Bourassa}, \binits{A.}},
    \bauthor{\bsnm{Heras}, \binits{F.J.H.}}, \betal:
    \batitle{Scaling and logic in the colour code on a superconducting quantum processor}.
    \bjtitle{Nature}
    (\byear{2025})
    \doiurl{10.1038/s41586-025-09061-4}
    \end{barticle}
    \endbibitem
    
    \bibitem[\protect\citeauthoryear{Arute et~al.}{2019}]{arute2019quantum}
    \begin{barticle}
    \bauthor{\bsnm{Arute}, \binits{F.}},
    \bauthor{\bsnm{Arya}, \binits{K.}},
    \bauthor{\bsnm{Babbush}, \binits{R.}},
    \bauthor{\bsnm{Bacon}, \binits{D.}},
    \bauthor{\bsnm{Bardin}, \binits{J.C.}},
    \bauthor{\bsnm{Barends}, \binits{R.}},
    \bauthor{\bsnm{Biswas}, \binits{R.}},
    \bauthor{\bsnm{Boixo}, \binits{S.}},
    \bauthor{\bsnm{Brandao}, \binits{F.G.}},
    \bauthor{\bsnm{Buell}, \binits{D.A.}}, \betal:
    \batitle{Quantum supremacy using a programmable superconducting processor}.
    \bjtitle{Nature}
    \bvolume{574}(\bissue{7779}),
    \bfpage{505}--\blpage{510}
    (\byear{2019})
    \end{barticle}
    \endbibitem
    
    \bibitem[\protect\citeauthoryear{Zhong et~al.}{2020}]{zhong2020quantum}
    \begin{barticle}
    \bauthor{\bsnm{Zhong}, \binits{H.-S.}},
    \bauthor{\bsnm{Wang}, \binits{H.}},
    \bauthor{\bsnm{Deng}, \binits{Y.-H.}},
    \bauthor{\bsnm{Chen}, \binits{M.-C.}},
    \bauthor{\bsnm{Peng}, \binits{L.-C.}},
    \bauthor{\bsnm{Luo}, \binits{Y.-H.}},
    \bauthor{\bsnm{Qin}, \binits{J.}},
    \bauthor{\bsnm{Wu}, \binits{D.}},
    \bauthor{\bsnm{Ding}, \binits{X.}},
    \bauthor{\bsnm{Hu}, \binits{Y.}}, \betal:
    \batitle{Quantum computational advantage using photons}.
    \bjtitle{Science}
    \bvolume{370}(\bissue{6523}),
    \bfpage{1460}--\blpage{1463}
    (\byear{2020})
    \end{barticle}
    \endbibitem
    
    \bibitem[\protect\citeauthoryear{Daley et~al.}{2022}]{daley2022practical}
    \begin{barticle}
    \bauthor{\bsnm{Daley}, \binits{A.J.}},
    \bauthor{\bsnm{Bloch}, \binits{I.}},
    \bauthor{\bsnm{Kokail}, \binits{C.}},
    \bauthor{\bsnm{Flannigan}, \binits{S.}},
    \bauthor{\bsnm{Pearson}, \binits{N.}},
    \bauthor{\bsnm{Troyer}, \binits{M.}},
    \bauthor{\bsnm{Zoller}, \binits{P.}}:
    \batitle{Practical quantum advantage in quantum simulation}.
    \bjtitle{Nature}
    \bvolume{607}(\bissue{7920}),
    \bfpage{667}--\blpage{676}
    (\byear{2022})
    \end{barticle}
    \endbibitem
    
    \bibitem[\protect\citeauthoryear{Lanes et~al.}{2025}]{lanes2025framework}
    \begin{botherref}
    \oauthor{\bsnm{Lanes}, \binits{O.}},
    \oauthor{\bsnm{Beji}, \binits{M.}},
    \oauthor{\bsnm{Corcoles}, \binits{A.D.}},
    \oauthor{\bsnm{Dalyac}, \binits{C.}},
    \oauthor{\bsnm{Gambetta}, \binits{J.M.}},
    \oauthor{\bsnm{Henriet}, \binits{L.}},
    \oauthor{\bsnm{Javadi-Abhari}, \binits{A.}},
    \oauthor{\bsnm{Kandala}, \binits{A.}},
    \oauthor{\bsnm{Mezzacapo}, \binits{A.}},
    \oauthor{\bsnm{Porter}, \binits{C.}}, et al.:
    A framework for quantum advantage.
    arXiv preprint arXiv:2506.20658
    (2025)
    \end{botherref}
    \endbibitem
    
    \bibitem[\protect\citeauthoryear{Kim et~al.}{2023}]{kim2023evidence}
    \begin{barticle}
    \bauthor{\bsnm{Kim}, \binits{Y.}},
    \bauthor{\bsnm{Eddins}, \binits{A.}},
    \bauthor{\bsnm{Anand}, \binits{S.}},
    \bauthor{\bsnm{Wei}, \binits{K.X.}},
    \bauthor{\bsnm{Van Den~Berg}, \binits{E.}},
    \bauthor{\bsnm{Rosenblatt}, \binits{S.}},
    \bauthor{\bsnm{Nayfeh}, \binits{H.}},
    \bauthor{\bsnm{Wu}, \binits{Y.}},
    \bauthor{\bsnm{Zaletel}, \binits{M.}},
    \bauthor{\bsnm{Temme}, \binits{K.}}, \betal:
    \batitle{Evidence for the utility of quantum computing before fault tolerance}.
    \bjtitle{Nature}
    \bvolume{618}(\bissue{7965}),
    \bfpage{500}--\blpage{505}
    (\byear{2023})
    \end{barticle}
    \endbibitem
    
    \bibitem[\protect\citeauthoryear{King et~al.}{2025}]{king2025beyond}
    \begin{barticle}
    \bauthor{\bsnm{King}, \binits{A.D.}},
    \bauthor{\bsnm{Nocera}, \binits{A.}},
    \bauthor{\bsnm{Rams}, \binits{M.M.}},
    \bauthor{\bsnm{Dziarmaga}, \binits{J.}},
    \bauthor{\bsnm{Wiersema}, \binits{R.}},
    \bauthor{\bsnm{Bernoudy}, \binits{W.}},
    \bauthor{\bsnm{Raymond}, \binits{J.}},
    \bauthor{\bsnm{Kaushal}, \binits{N.}},
    \bauthor{\bsnm{Heinsdorf}, \binits{N.}},
    \bauthor{\bsnm{Harris}, \binits{R.}}, \betal:
    \batitle{Beyond-classical computation in quantum simulation}.
    \bjtitle{Science}
    \bvolume{388}(\bissue{6743}),
    \bfpage{199}--\blpage{204}
    (\byear{2025})
    \end{barticle}
    \endbibitem
    
    \bibitem[\protect\citeauthoryear{Tindall et~al.}{2024}]{tindall2024efficient}
    \begin{barticle}
    \bauthor{\bsnm{Tindall}, \binits{J.}},
    \bauthor{\bsnm{Fishman}, \binits{M.}},
    \bauthor{\bsnm{Stoudenmire}, \binits{E.M.}},
    \bauthor{\bsnm{Sels}, \binits{D.}}:
    \batitle{Efficient tensor network simulation of ibm’s eagle kicked ising experiment}.
    \bjtitle{Prx quantum}
    \bvolume{5}(\bissue{1}),
    \bfpage{010308}
    (\byear{2024})
    \end{barticle}
    \endbibitem
    
    \bibitem[\protect\citeauthoryear{Tindall et~al.}{2025}]{tindall2025dynamics}
    \begin{botherref}
    \oauthor{\bsnm{Tindall}, \binits{J.}},
    \oauthor{\bsnm{Mello}, \binits{A.}},
    \oauthor{\bsnm{Fishman}, \binits{M.}},
    \oauthor{\bsnm{Stoudenmire}, \binits{M.}},
    \oauthor{\bsnm{Sels}, \binits{D.}}:
    Dynamics of disordered quantum systems with two-and three-dimensional tensor networks.
    arXiv preprint arXiv:2503.05693
    (2025)
    \end{botherref}
    \endbibitem
    
    \bibitem[\protect\citeauthoryear{Mauron and Carleo}{2025}]{mauron2025challenging}
    \begin{botherref}
    \oauthor{\bsnm{Mauron}, \binits{L.}},
    \oauthor{\bsnm{Carleo}, \binits{G.}}:
    Challenging the quantum advantage frontier with large-scale classical simulations of annealing dynamics.
    arXiv preprint arXiv:2503.08247
    (2025)
    \end{botherref}
    \endbibitem
    
    \bibitem[\protect\citeauthoryear{Campbell}{2021}]{campbell2021early}
    \begin{barticle}
    \bauthor{\bsnm{Campbell}, \binits{E.T.}}:
    \batitle{Early fault-tolerant simulations of the hubbard model}.
    \bjtitle{Quantum Science and Technology}
    \bvolume{7}(\bissue{1}),
    \bfpage{015007}
    (\byear{2021})
    \end{barticle}
    \endbibitem
    
    \bibitem[\protect\citeauthoryear{Katabarwa et~al.}{2024}]{katabarwa2024early}
    \begin{barticle}
    \bauthor{\bsnm{Katabarwa}, \binits{A.}},
    \bauthor{\bsnm{Gratsea}, \binits{K.}},
    \bauthor{\bsnm{Caesura}, \binits{A.}},
    \bauthor{\bsnm{Johnson}, \binits{P.D.}}:
    \batitle{Early fault-tolerant quantum computing}.
    \bjtitle{PRX Quantum}
    \bvolume{5}(\bissue{2}),
    \bfpage{020101}
    (\byear{2024})
    \end{barticle}
    \endbibitem
    
    \bibitem[\protect\citeauthoryear{Preskill}{2025}]{preskill2025beyond}
    \begin{botherref}
    \oauthor{\bsnm{Preskill}, \binits{J.}}:
    Beyond nisq: The megaquop machine.
    ACM New York, NY
    (2025)
    \end{botherref}
    \endbibitem
    
    \bibitem[\protect\citeauthoryear{Lee et~al.}{2021}]{lee2021even}
    \begin{barticle}
    \bauthor{\bsnm{Lee}, \binits{J.}},
    \bauthor{\bsnm{Berry}, \binits{D.W.}},
    \bauthor{\bsnm{Gidney}, \binits{C.}},
    \bauthor{\bsnm{Huggins}, \binits{W.J.}},
    \bauthor{\bsnm{McClean}, \binits{J.R.}},
    \bauthor{\bsnm{Wiebe}, \binits{N.}},
    \bauthor{\bsnm{Babbush}, \binits{R.}}:
    \batitle{Even more efficient quantum computations of chemistry through tensor hypercontraction}.
    \bjtitle{PRX Quantum}
    \bvolume{2}(\bissue{3}),
    \bfpage{030305}
    (\byear{2021})
    \end{barticle}
    \endbibitem
    
    \bibitem[\protect\citeauthoryear{Caesura et~al.}{2025}]{caesura2025faster}
    \begin{botherref}
    \oauthor{\bsnm{Caesura}, \binits{A.}},
    \oauthor{\bsnm{Cortes}, \binits{C.L.}},
    \oauthor{\bsnm{Pol}, \binits{W.}},
    \oauthor{\bsnm{Sim}, \binits{S.}},
    \oauthor{\bsnm{Steudtner}, \binits{M.}},
    \oauthor{\bsnm{Anselmetti}, \binits{G.-L.R.}},
    \oauthor{\bsnm{Degroote}, \binits{M.}},
    \oauthor{\bsnm{Moll}, \binits{N.}},
    \oauthor{\bsnm{Santagati}, \binits{R.}},
    \oauthor{\bsnm{Streif}, \binits{M.}}, et al.:
    Faster quantum chemistry simulations on a quantum computer with improved tensor factorization and active volume compilation.
    arXiv preprint arXiv:2501.06165
    (2025)
    \end{botherref}
    \endbibitem
    
    \bibitem[\protect\citeauthoryear{Gidney}{2025}]{gidney2025factor}
    \begin{botherref}
    \oauthor{\bsnm{Gidney}, \binits{C.}}:
    How to factor 2048 bit rsa integers with less than a million noisy qubits.
    arXiv preprint arXiv:2505.15917
    (2025)
    \end{botherref}
    \endbibitem
    
    \bibitem[\protect\citeauthoryear{Iyer and Poulin}{2018}]{iyer2018small}
    \begin{barticle}
    \bauthor{\bsnm{Iyer}, \binits{P.}},
    \bauthor{\bsnm{Poulin}, \binits{D.}}:
    \batitle{A small quantum computer is needed to optimize fault-tolerant protocols}.
    \bjtitle{Quantum Science and Technology}
    \bvolume{3}(\bissue{3}),
    \bfpage{030504}
    (\byear{2018})
    \end{barticle}
    \endbibitem
    
    \bibitem[\protect\citeauthoryear{Kyaw et~al.}{2021}]{kyaw2021quantum}
    \begin{barticle}
    \bauthor{\bsnm{Kyaw}, \binits{T.H.}},
    \bauthor{\bsnm{Menke}, \binits{T.}},
    \bauthor{\bsnm{Sim}, \binits{S.}},
    \bauthor{\bsnm{Anand}, \binits{A.}},
    \bauthor{\bsnm{Sawaya}, \binits{N.P.}},
    \bauthor{\bsnm{Oliver}, \binits{W.D.}},
    \bauthor{\bsnm{Guerreschi}, \binits{G.G.}},
    \bauthor{\bsnm{Aspuru-Guzik}, \binits{A.}}:
    \batitle{Quantum computer-aided design: digital quantum simulation of quantum processors}.
    \bjtitle{Physical Review Applied}
    \bvolume{16}(\bissue{4}),
    \bfpage{044042}
    (\byear{2021})
    \end{barticle}
    \endbibitem
    
    \bibitem[\protect\citeauthoryear{Lie}{1893}]{lie1893theorie}
    \begin{bbook}
    \bauthor{\bsnm{Lie}, \binits{S.}}:
    \bbtitle{Theorie der Transformationsgruppen}
    vol. \bseriesno{3}.
    \bpublisher{Teubner},
    \blocation{Leipzig, Germany}
    (\byear{1893})
    \end{bbook}
    \endbibitem
    
    \bibitem[\protect\citeauthoryear{Trotter}{1958}]{trotter1958approximation}
    \begin{botherref}
    \oauthor{\bsnm{Trotter}, \binits{H.F.}}:
    Approximation of semi-groups of operators.
    Pacific Journal of Mathematics
    \textbf{8}(4)
    (1958)
    \end{botherref}
    \endbibitem
    
    \bibitem[\protect\citeauthoryear{Trotter}{1959}]{trotter1959product}
    \begin{barticle}
    \bauthor{\bsnm{Trotter}, \binits{H.F.}}:
    \batitle{On the product of semi-groups of operators}.
    \bjtitle{Proceedings of the American Mathematical Society}
    \bvolume{10}(\bissue{4}),
    \bfpage{545}--\blpage{551}
    (\byear{1959})
    \end{barticle}
    \endbibitem
    
    \bibitem[\protect\citeauthoryear{Suzuki}{1976}]{suzuki1976generalized}
    \begin{barticle}
    \bauthor{\bsnm{Suzuki}, \binits{M.}}:
    \batitle{Generalized trotter's formula and systematic approximants of exponential operators and inner derivations with applications to many-body problems}.
    \bjtitle{Communications in Mathematical Physics}
    \bvolume{51}(\bissue{2}),
    \bfpage{183}--\blpage{190}
    (\byear{1976})
    \end{barticle}
    \endbibitem
    
    \bibitem[\protect\citeauthoryear{Suzuki}{1985}]{suzuki1985decomposition}
    \begin{barticle}
    \bauthor{\bsnm{Suzuki}, \binits{M.}}:
    \batitle{Decomposition formulas of exponential operators and lie exponentials with some applications to quantum mechanics and statistical physics}.
    \bjtitle{Journal of mathematical physics}
    \bvolume{26}(\bissue{4}),
    \bfpage{601}--\blpage{612}
    (\byear{1985})
    \end{barticle}
    \endbibitem
    
    \bibitem[\protect\citeauthoryear{Suzuki}{1986}]{suzuki1986quantum}
    \begin{barticle}
    \bauthor{\bsnm{Suzuki}, \binits{M.}}:
    \batitle{Quantum statistical monte carlo methods and applications to spin systems}.
    \bjtitle{Journal of Statistical Physics}
    \bvolume{43},
    \bfpage{883}--\blpage{909}
    (\byear{1986})
    \end{barticle}
    \endbibitem
    
    \bibitem[\protect\citeauthoryear{Suzuki}{2012}]{suzuki2012quantum}
    \begin{bbook}
    \bauthor{\bsnm{Suzuki}, \binits{M.}}:
    \bbtitle{Quantum Monte Carlo Methods in Equilibrium and Nonequilibrium Systems: Proceedings of the Ninth Taniguchi International Symposium, Susono, Japan, November 14--18, 1986}
    vol. \bseriesno{74}.
    \bpublisher{Springer},
    \blocation{Berlin/Heidelberg, Germany}
    (\byear{2012})
    \end{bbook}
    \endbibitem
    
    \bibitem[\protect\citeauthoryear{Suzuki}{1990}]{suzuki1990fractal}
    \begin{barticle}
    \bauthor{\bsnm{Suzuki}, \binits{M.}}:
    \batitle{Fractal decomposition of exponential operators with applications to many-body theories and monte carlo simulations}.
    \bjtitle{Physics Letters A}
    \bvolume{146}(\bissue{6}),
    \bfpage{319}--\blpage{323}
    (\byear{1990})
    \end{barticle}
    \endbibitem
    
    \bibitem[\protect\citeauthoryear{Hatano and Suzuki}{2005}]{hatano2005finding}
    \begin{bchapter}
    \bauthor{\bsnm{Hatano}, \binits{N.}},
    \bauthor{\bsnm{Suzuki}, \binits{M.}}:
    \bctitle{Finding exponential product formulas of higher orders}.
    In: \bbtitle{Quantum Annealing and Other Optimization Methods},
    pp. \bfpage{37}--\blpage{68}.
    \bpublisher{Springer},
    \blocation{Berlin/Heidelberg, Germany}
    (\byear{2005})
    \end{bchapter}
    \endbibitem
    
    \bibitem[\protect\citeauthoryear{Kivlichan et~al.}{2020}]{kivlichan2020improved}
    \begin{barticle}
    \bauthor{\bsnm{Kivlichan}, \binits{I.D.}},
    \bauthor{\bsnm{Gidney}, \binits{C.}},
    \bauthor{\bsnm{Berry}, \binits{D.W.}},
    \bauthor{\bsnm{Wiebe}, \binits{N.}},
    \bauthor{\bsnm{McClean}, \binits{J.}},
    \bauthor{\bsnm{Sun}, \binits{W.}},
    \bauthor{\bsnm{Jiang}, \binits{Z.}},
    \bauthor{\bsnm{Rubin}, \binits{N.}},
    \bauthor{\bsnm{Fowler}, \binits{A.}},
    \bauthor{\bsnm{Aspuru-Guzik}, \binits{A.}}, \betal:
    \batitle{Improved fault-tolerant quantum simulation of condensed-phase correlated electrons via trotterization}.
    \bjtitle{Quantum}
    \bvolume{4},
    \bfpage{296}
    (\byear{2020})
    \end{barticle}
    \endbibitem
    
    \bibitem[\protect\citeauthoryear{Childs et~al.}{2018}]{childs2018toward}
    \begin{barticle}
    \bauthor{\bsnm{Childs}, \binits{A.M.}},
    \bauthor{\bsnm{Maslov}, \binits{D.}},
    \bauthor{\bsnm{Nam}, \binits{Y.}},
    \bauthor{\bsnm{Ross}, \binits{N.J.}},
    \bauthor{\bsnm{Su}, \binits{Y.}}:
    \batitle{Toward the first quantum simulation with quantum speedup}.
    \bjtitle{Proceedings of the National Academy of Sciences}
    \bvolume{115}(\bissue{38}),
    \bfpage{9456}--\blpage{9461}
    (\byear{2018})
    \end{barticle}
    \endbibitem
    
    \bibitem[\protect\citeauthoryear{Childs et~al.}{2021}]{childs2021theory}
    \begin{barticle}
    \bauthor{\bsnm{Childs}, \binits{A.M.}},
    \bauthor{\bsnm{Su}, \binits{Y.}},
    \bauthor{\bsnm{Tran}, \binits{M.C.}},
    \bauthor{\bsnm{Wiebe}, \binits{N.}},
    \bauthor{\bsnm{Zhu}, \binits{S.}}:
    \batitle{Theory of trotter error with commutator scaling}.
    \bjtitle{Physical Review X}
    \bvolume{11}(\bissue{1}),
    \bfpage{011020}
    (\byear{2021})
    \end{barticle}
    \endbibitem
    
    \bibitem[\protect\citeauthoryear{Yi and Crosson}{2022}]{yi2022spectral}
    \begin{barticle}
    \bauthor{\bsnm{Yi}, \binits{C.}},
    \bauthor{\bsnm{Crosson}, \binits{E.}}:
    \batitle{Spectral analysis of product formulas for quantum simulation}.
    \bjtitle{npj Quantum Information}
    \bvolume{8}(\bissue{1}),
    \bfpage{37}
    (\byear{2022})
    \end{barticle}
    \endbibitem
    
    \bibitem[\protect\citeauthoryear{Cleve et~al.}{1998}]{cleve1998quantum}
    \begin{barticle}
    \bauthor{\bsnm{Cleve}, \binits{R.}},
    \bauthor{\bsnm{Ekert}, \binits{A.}},
    \bauthor{\bsnm{Macchiavello}, \binits{C.}},
    \bauthor{\bsnm{Mosca}, \binits{M.}}:
    \batitle{Quantum algorithms revisited}.
    \bjtitle{Proceedings of the Royal Society of London. Series A: Mathematical, Physical and Engineering Sciences}
    \bvolume{454}(\bissue{1969}),
    \bfpage{339}--\blpage{354}
    (\byear{1998})
    \end{barticle}
    \endbibitem
    
    \bibitem[\protect\citeauthoryear{Brassard et~al.}{2000}]{brassard2000quantum}
    \begin{botherref}
    \oauthor{\bsnm{Brassard}, \binits{G.}},
    \oauthor{\bsnm{Hoyer}, \binits{P.}},
    \oauthor{\bsnm{Mosca}, \binits{M.}},
    \oauthor{\bsnm{Tapp}, \binits{A.}}:
    Quantum amplitude amplification and estimation (2000).
    arXiv preprint quant-ph/0005055
    (2000)
    \end{botherref}
    \endbibitem
    
    \bibitem[\protect\citeauthoryear{Grinko et~al.}{2021}]{grinko2021iterative}
    \begin{barticle}
    \bauthor{\bsnm{Grinko}, \binits{D.}},
    \bauthor{\bsnm{Gacon}, \binits{J.}},
    \bauthor{\bsnm{Zoufal}, \binits{C.}},
    \bauthor{\bsnm{Woerner}, \binits{S.}}:
    \batitle{Iterative quantum amplitude estimation}.
    \bjtitle{npj Quantum Information}
    \bvolume{7}(\bissue{1}),
    \bfpage{52}
    (\byear{2021})
    \end{barticle}
    \endbibitem
    
    \bibitem[\protect\citeauthoryear{Suzuki et~al.}{2020}]{suzuki2020amplitude}
    \begin{barticle}
    \bauthor{\bsnm{Suzuki}, \binits{Y.}},
    \bauthor{\bsnm{Uno}, \binits{S.}},
    \bauthor{\bsnm{Raymond}, \binits{R.}},
    \bauthor{\bsnm{Tanaka}, \binits{T.}},
    \bauthor{\bsnm{Onodera}, \binits{T.}},
    \bauthor{\bsnm{Yamamoto}, \binits{N.}}:
    \batitle{Amplitude estimation without phase estimation}.
    \bjtitle{Quantum Information Processing}
    \bvolume{19},
    \bfpage{1}--\blpage{17}
    (\byear{2020})
    \end{barticle}
    \endbibitem
    
    \bibitem[\protect\citeauthoryear{Gily{\'e}n et~al.}{2019}]{gilyen2019quantum}
    \begin{bchapter}
    \bauthor{\bsnm{Gily{\'e}n}, \binits{A.}},
    \bauthor{\bsnm{Su}, \binits{Y.}},
    \bauthor{\bsnm{Low}, \binits{G.H.}},
    \bauthor{\bsnm{Wiebe}, \binits{N.}}:
    \bctitle{Quantum singular value transformation and beyond: exponential improvements for quantum matrix arithmetics}.
    In: \bbtitle{Proceedings of the 51st Annual ACM SIGACT Symposium on Theory of Computing},
    pp. \bfpage{193}--\blpage{204}
    (\byear{2019})
    \end{bchapter}
    \endbibitem
    
    \bibitem[\protect\citeauthoryear{Poulin et~al.}{2015}]{poulin2014trotter}
    \begin{barticle}
    \bauthor{\bsnm{Poulin}, \binits{D.}},
    \bauthor{\bsnm{Hastings}, \binits{M.B.}},
    \bauthor{\bsnm{Wecker}, \binits{D.}},
    \bauthor{\bsnm{Wiebe}, \binits{N.}},
    \bauthor{\bsnm{Doberty}, \binits{A.C.}},
    \bauthor{\bsnm{Troyer}, \binits{M.}}:
    \batitle{The trotter step size required for accurate quantum simulation of quantum chemistry}.
    \bjtitle{Quantum Info. Comput.}
    \bvolume{15}(\bissue{5–6}),
    \bfpage{361}--\blpage{384}
    (\byear{2015})
    \end{barticle}
    \endbibitem
    
    \bibitem[\protect\citeauthoryear{Tang et~al.}{2021}]{tang2021qubit}
    \begin{barticle}
    \bauthor{\bsnm{Tang}, \binits{H.L.}},
    \bauthor{\bsnm{Shkolnikov}, \binits{V.}},
    \bauthor{\bsnm{Barron}, \binits{G.S.}},
    \bauthor{\bsnm{Grimsley}, \binits{H.R.}},
    \bauthor{\bsnm{Mayhall}, \binits{N.J.}},
    \bauthor{\bsnm{Barnes}, \binits{E.}},
    \bauthor{\bsnm{Economou}, \binits{S.E.}}:
    \batitle{qubit-adapt-vqe: An adaptive algorithm for constructing hardware-efficient ans{\"a}tze on a quantum processor}.
    \bjtitle{PRX Quantum}
    \bvolume{2}(\bissue{2}),
    \bfpage{020310}
    (\byear{2021})
    \end{barticle}
    \endbibitem
    
    \bibitem[\protect\citeauthoryear{Babbush et~al.}{2014}]{babbush2014adiabatic}
    \begin{barticle}
    \bauthor{\bsnm{Babbush}, \binits{R.}},
    \bauthor{\bsnm{Love}, \binits{P.J.}},
    \bauthor{\bsnm{Aspuru-Guzik}, \binits{A.}}:
    \batitle{Adiabatic quantum simulation of quantum chemistry}.
    \bjtitle{Scientific reports}
    \bvolume{4}(\bissue{1}),
    \bfpage{6603}
    (\byear{2014})
    \end{barticle}
    \endbibitem
    
    \bibitem[\protect\citeauthoryear{Babbush et~al.}{2015}]{babbush2015chemical}
    \begin{barticle}
    \bauthor{\bsnm{Babbush}, \binits{R.}},
    \bauthor{\bsnm{McClean}, \binits{J.}},
    \bauthor{\bsnm{Wecker}, \binits{D.}},
    \bauthor{\bsnm{Aspuru-Guzik}, \binits{A.}},
    \bauthor{\bsnm{Wiebe}, \binits{N.}}:
    \batitle{Chemical basis of trotter-suzuki errors in quantum chemistry simulation}.
    \bjtitle{Physical Review A}
    \bvolume{91}(\bissue{2}),
    \bfpage{022311}
    (\byear{2015})
    \end{barticle}
    \endbibitem
    
    \bibitem[\protect\citeauthoryear{Hastings et~al.}{2015}]{hastings2014improving}
    \begin{barticle}
    \bauthor{\bsnm{Hastings}, \binits{M.B.}},
    \bauthor{\bsnm{Wecker}, \binits{D.}},
    \bauthor{\bsnm{Bauer}, \binits{B.}},
    \bauthor{\bsnm{Troyer}, \binits{M.}}:
    \batitle{Improving quantum algorithms for quantum chemistry}.
    \bjtitle{Quantum Info. Comput.}
    \bvolume{15}(\bissue{1–2}),
    \bfpage{1}--\blpage{21}
    (\byear{2015})
    \end{barticle}
    \endbibitem
    
    \bibitem[\protect\citeauthoryear{Wecker et~al.}{2014}]{wecker2014gate}
    \begin{barticle}
    \bauthor{\bsnm{Wecker}, \binits{D.}},
    \bauthor{\bsnm{Bauer}, \binits{B.}},
    \bauthor{\bsnm{Clark}, \binits{B.K.}},
    \bauthor{\bsnm{Hastings}, \binits{M.B.}},
    \bauthor{\bsnm{Troyer}, \binits{M.}}:
    \batitle{Gate-count estimates for performing quantum chemistry on small quantum computers}.
    \bjtitle{Phys. Rev. A}
    \bvolume{90},
    \bfpage{022305}
    (\byear{2014})
    \doiurl{10.1103/PhysRevA.90.022305}
    \end{barticle}
    \endbibitem
    
    \bibitem[\protect\citeauthoryear{Raeisi et~al.}{2012}]{raeisi2012quantum}
    \begin{barticle}
    \bauthor{\bsnm{Raeisi}, \binits{S.}},
    \bauthor{\bsnm{Wiebe}, \binits{N.}},
    \bauthor{\bsnm{Sanders}, \binits{B.C.}}:
    \batitle{Quantum-circuit design for efficient simulations of many-body quantum dynamics}.
    \bjtitle{New Journal of Physics}
    \bvolume{14}(\bissue{10}),
    \bfpage{103017}
    (\byear{2012})
    \end{barticle}
    \endbibitem
    
    \bibitem[\protect\citeauthoryear{Nielsen and Chuang}{2001}]{nielsen2001quantum}
    \begin{bbook}
    \bauthor{\bsnm{Nielsen}, \binits{M.A.}},
    \bauthor{\bsnm{Chuang}, \binits{I.L.}}:
    \bbtitle{Quantum Computation and Quantum Information}
    vol. \bseriesno{2}.
    \bpublisher{Cambridge University Press},
    \blocation{Cambridge, UK}
    (\byear{2001})
    \end{bbook}
    \endbibitem
    
    \bibitem[\protect\citeauthoryear{Childs and Kothari}{2010}]{childs2009limitations}
    \begin{barticle}
    \bauthor{\bsnm{Childs}, \binits{A.M.}},
    \bauthor{\bsnm{Kothari}, \binits{R.}}:
    \batitle{Limitations on the simulation of non-sparse hamiltonians}.
    \bjtitle{Quantum Info. Comput.}
    \bvolume{10}(\bissue{7}),
    \bfpage{669}--\blpage{684}
    (\byear{2010})
    \end{barticle}
    \endbibitem
    
    \bibitem[\protect\citeauthoryear{Hempel et~al.}{2018}]{hempel2018quantum}
    \begin{barticle}
    \bauthor{\bsnm{Hempel}, \binits{C.}},
    \bauthor{\bsnm{Maier}, \binits{C.}},
    \bauthor{\bsnm{Romero}, \binits{J.}},
    \bauthor{\bsnm{McClean}, \binits{J.}},
    \bauthor{\bsnm{Monz}, \binits{T.}},
    \bauthor{\bsnm{Shen}, \binits{H.}},
    \bauthor{\bsnm{Jurcevic}, \binits{P.}},
    \bauthor{\bsnm{Lanyon}, \binits{B.P.}},
    \bauthor{\bsnm{Love}, \binits{P.}},
    \bauthor{\bsnm{Babbush}, \binits{R.}}, \betal:
    \batitle{Quantum chemistry calculations on a trapped-ion quantum simulator}.
    \bjtitle{Physical Review X}
    \bvolume{8}(\bissue{3}),
    \bfpage{031022}
    (\byear{2018})
    \end{barticle}
    \endbibitem
    
    \bibitem[\protect\citeauthoryear{Tranter et~al.}{2019}]{tranter2019ordering}
    \begin{barticle}
    \bauthor{\bsnm{Tranter}, \binits{A.}},
    \bauthor{\bsnm{Love}, \binits{P.J.}},
    \bauthor{\bsnm{Mintert}, \binits{F.}},
    \bauthor{\bsnm{Wiebe}, \binits{N.}},
    \bauthor{\bsnm{Coveney}, \binits{P.V.}}:
    \batitle{Ordering of trotterization: Impact on errors in quantum simulation of electronic structure}.
    \bjtitle{Entropy}
    \bvolume{21}(\bissue{12}),
    \bfpage{1218}
    (\byear{2019})
    \end{barticle}
    \endbibitem
    
    \bibitem[\protect\citeauthoryear{Burgarth et~al.}{2024}]{burgarth2024strong}
    \begin{barticle}
    \bauthor{\bsnm{Burgarth}, \binits{D.}},
    \bauthor{\bsnm{Facchi}, \binits{P.}},
    \bauthor{\bsnm{Hahn}, \binits{A.}},
    \bauthor{\bsnm{Johnsson}, \binits{M.}},
    \bauthor{\bsnm{Yuasa}, \binits{K.}}:
    \batitle{Strong error bounds for trotter and strang-splittings and their implications for quantum chemistry}.
    \bjtitle{Physical Review Research}
    \bvolume{6}(\bissue{4}),
    \bfpage{043155}
    (\byear{2024})
    \end{barticle}
    \endbibitem
    
    \bibitem[\protect\citeauthoryear{Higgins et~al.}{2007}]{higgins2007entanglement}
    \begin{barticle}
    \bauthor{\bsnm{Higgins}, \binits{B.L.}},
    \bauthor{\bsnm{Berry}, \binits{D.W.}},
    \bauthor{\bsnm{Bartlett}, \binits{S.D.}},
    \bauthor{\bsnm{Wiseman}, \binits{H.M.}},
    \bauthor{\bsnm{Pryde}, \binits{G.J.}}:
    \batitle{Entanglement-free heisenberg-limited phase estimation}.
    \bjtitle{Nature}
    \bvolume{450}(\bissue{7168}),
    \bfpage{393}--\blpage{396}
    (\byear{2007})
    \end{barticle}
    \endbibitem
    
    \bibitem[\protect\citeauthoryear{Berry et~al.}{2009}]{berry2009}
    \begin{barticle}
    \bauthor{\bsnm{Berry}, \binits{D.W.}},
    \bauthor{\bsnm{Higgins}, \binits{B.L.}},
    \bauthor{\bsnm{Bartlett}, \binits{S.D.}},
    \bauthor{\bsnm{Mitchell}, \binits{M.W.}},
    \bauthor{\bsnm{Pryde}, \binits{G.J.}},
    \bauthor{\bsnm{Wiseman}, \binits{H.M.}}:
    \batitle{How to perform the most accurate possible phase measurements}.
    \bjtitle{Phys. Rev. A}
    \bvolume{80},
    \bfpage{052114}
    (\byear{2009})
    \doiurl{10.1103/PhysRevA.80.052114}
    \end{barticle}
    \endbibitem
    
    \bibitem[\protect\citeauthoryear{Bhatia and Davis}{1984}]{bhatia1984bound}
    \begin{barticle}
    \bauthor{\bsnm{Bhatia}, \binits{R.}},
    \bauthor{\bsnm{Davis}, \binits{C.}}:
    \batitle{A bound for the spectral variation of a unitary operator}.
    \bjtitle{Linear and Multilinear Algebra}
    \bvolume{15}(\bissue{1}),
    \bfpage{71}--\blpage{76}
    (\byear{1984})
    \end{barticle}
    \endbibitem
    
    \bibitem[\protect\citeauthoryear{Baker}{1905}]{baker1905alternants}
    \begin{barticle}
    \bauthor{\bsnm{Baker}, \binits{H.F.}}:
    \batitle{Alternants and continuous groups}.
    \bjtitle{Proceedings of the London Mathematical Society}
    \bvolume{2}(\bissue{1}),
    \bfpage{24}--\blpage{47}
    (\byear{1905})
    \end{barticle}
    \endbibitem
    
    \bibitem[\protect\citeauthoryear{Hausdorff}{1906}]{hausdorff1906symbolische}
    \begin{barticle}
    \bauthor{\bsnm{Hausdorff}, \binits{F.}}:
    \batitle{Die symbolische exponentialformel in der gruppentheorie}.
    \bjtitle{Ber. Verh. Kgl. S{\~A}¤ chs. Ges. Wiss. Leipzig., Math.-phys. Kl.}
    \bvolume{58},
    \bfpage{19}--\blpage{48}
    (\byear{1906})
    \end{barticle}
    \endbibitem
    
    \bibitem[\protect\citeauthoryear{Jordan and Wigner}{1928}]{jordan1928paulische}
    \begin{botherref}
    \oauthor{\bsnm{Jordan}, \binits{P.}},
    \oauthor{\bsnm{Wigner}, \binits{E.}}:
    {\"u}ber das Paulische {\"a}quivalenzverbot. Z Phys 47: 631
    (1928)
    \end{botherref}
    \endbibitem
    
    \bibitem[\protect\citeauthoryear{Bravyi et~al.}{2017}]{bravyi2017tapering}
    \begin{botherref}
    \oauthor{\bsnm{Bravyi}, \binits{S.}},
    \oauthor{\bsnm{Gambetta}, \binits{J.M.}},
    \oauthor{\bsnm{Mezzacapo}, \binits{A.}},
    \oauthor{\bsnm{Temme}, \binits{K.}}:
    Tapering off qubits to simulate fermionic hamiltonians.
    arXiv preprint arXiv:1701.08213
    (2017)
    \end{botherref}
    \endbibitem
    
    \bibitem[\protect\citeauthoryear{Kirby et~al.}{2021}]{kirby2021contextual}
    \begin{barticle}
    \bauthor{\bsnm{Kirby}, \binits{W.M.}},
    \bauthor{\bsnm{Tranter}, \binits{A.}},
    \bauthor{\bsnm{Love}, \binits{P.J.}}:
    \batitle{Contextual subspace variational quantum eigensolver}.
    \bjtitle{Quantum}
    \bvolume{5},
    \bfpage{456}
    (\byear{2021})
    \end{barticle}
    \endbibitem
    
    \bibitem[\protect\citeauthoryear{Weaving et~al.}{2023}]{weaving2023stabilizer}
    \begin{barticle}
    \bauthor{\bsnm{Weaving}, \binits{T.}},
    \bauthor{\bsnm{Ralli}, \binits{A.}},
    \bauthor{\bsnm{Kirby}, \binits{W.M.}},
    \bauthor{\bsnm{Tranter}, \binits{A.}},
    \bauthor{\bsnm{Love}, \binits{P.J.}},
    \bauthor{\bsnm{Coveney}, \binits{P.V.}}:
    \batitle{A stabilizer framework for the contextual subspace variational quantum eigensolver and the noncontextual projection ansatz}.
    \bjtitle{Journal of Chemical Theory and Computation}
    \bvolume{19}(\bissue{3}),
    \bfpage{808}--\blpage{821}
    (\byear{2023})
    \end{barticle}
    \endbibitem
    
    \bibitem[\protect\citeauthoryear{Ralli et~al.}{2023}]{ralli2023unitary}
    \begin{barticle}
    \bauthor{\bsnm{Ralli}, \binits{A.}},
    \bauthor{\bsnm{Weaving}, \binits{T.}},
    \bauthor{\bsnm{Tranter}, \binits{A.}},
    \bauthor{\bsnm{Kirby}, \binits{W.M.}},
    \bauthor{\bsnm{Love}, \binits{P.J.}},
    \bauthor{\bsnm{Coveney}, \binits{P.V.}}:
    \batitle{Unitary partitioning and the contextual subspace variational quantum eigensolver}.
    \bjtitle{Physical Review Research}
    \bvolume{5}(\bissue{1}),
    \bfpage{013095}
    (\byear{2023})
    \end{barticle}
    \endbibitem
    
    \bibitem[\protect\citeauthoryear{Weaving et~al.}{2025}]{weaving2023contextual}
    \begin{barticle}
    \bauthor{\bsnm{Weaving}, \binits{T.}},
    \bauthor{\bsnm{Ralli}, \binits{A.}},
    \bauthor{\bsnm{Love}, \binits{P.J.}},
    \bauthor{\bsnm{Succi}, \binits{S.}},
    \bauthor{\bsnm{Coveney}, \binits{P.V.}}:
    \batitle{Contextual subspace variational quantum eigensolver calculation of the dissociation curve of molecular nitrogen on a superconducting quantum computer}.
    \bjtitle{npj Quantum Information}
    \bvolume{11}(\bissue{1}),
    \bfpage{25}
    (\byear{2025})
    \end{barticle}
    \endbibitem
    
    \bibitem[\protect\citeauthoryear{Berry et~al.}{2025}]{google_mps_prep}
    \begin{barticle}
    \bauthor{\bsnm{Berry}, \binits{D.W.}},
    \bauthor{\bsnm{Tong}, \binits{Y.}},
    \bauthor{\bsnm{Khattar}, \binits{T.}},
    \bauthor{\bsnm{White}, \binits{A.}},
    \bauthor{\bsnm{Kim}, \binits{T.I.}},
    \bauthor{\bsnm{Low}, \binits{G.H.}},
    \bauthor{\bsnm{Boixo}, \binits{S.}},
    \bauthor{\bsnm{Ding}, \binits{Z.}},
    \bauthor{\bsnm{Lin}, \binits{L.}},
    \bauthor{\bsnm{Lee}, \binits{S.}},
    \bauthor{\bsnm{Chan}, \binits{G.K.-L.}},
    \bauthor{\bsnm{Babbush}, \binits{R.}},
    \bauthor{\bsnm{Rubin}, \binits{N.C.}}:
    \batitle{Rapid initial-state preparation for the quantum simulation of strongly correlated molecules}.
    \bjtitle{PRX Quantum}
    \bvolume{6},
    \bfpage{020327}
    (\byear{2025})
    \doiurl{10.1103/PRXQuantum.6.020327}
    \end{barticle}
    \endbibitem
    
    \bibitem[\protect\citeauthoryear{Farhi et~al.}{2000}]{farhi2000quantum}
    \begin{botherref}
    \oauthor{\bsnm{Farhi}, \binits{E.}},
    \oauthor{\bsnm{Goldstone}, \binits{J.}},
    \oauthor{\bsnm{Gutmann}, \binits{S.}},
    \oauthor{\bsnm{Sipser}, \binits{M.}}:
    Quantum computation by adiabatic evolution.
    arXiv preprint quant-ph/0001106
    (2000)
    \end{botherref}
    \endbibitem
    
    \bibitem[\protect\citeauthoryear{Albash and Lidar}{2018}]{albash2018adiabatic}
    \begin{barticle}
    \bauthor{\bsnm{Albash}, \binits{T.}},
    \bauthor{\bsnm{Lidar}, \binits{D.A.}}:
    \batitle{Adiabatic quantum computation}.
    \bjtitle{Reviews of Modern Physics}
    \bvolume{90}(\bissue{1}),
    \bfpage{015002}
    (\byear{2018})
    \end{barticle}
    \endbibitem
    
    \end{thebibliography}
\end{document}